\documentclass[11pt]{article}
\newtheorem{theorem}{Theorem}
\newtheorem{proposition}[theorem]{Proposition}
\newtheorem{definition}[theorem]{Definition}

\newtheorem{lemma}[theorem]{Lemma}
\usepackage{style}
\usepackage{geometry}      
\usepackage{amsmath}
\usepackage[pstricks1-10]{vaucanson-g}
\usepackage{graphicx}
\usepackage{amssymb}
\usepackage{epstopdf}
\usepackage{algorithm}
\usepackage{ALgo}

\title{\Large\bf Algorithms for Glushkov $\K$-graphs
}

\author{
Pascal {\sc
Caron}\thanks{Pascal.Caron@univ-rouen.fr}, Marianne
{\sc Flouret} \thanks{Marianne.Flouret@univ-lehavre.fr}
\\ LITIS,
Universit\'e de Rouen, 76801
Saint \'Etienne du Rouvray, France \\ LITIS, Universit\'e du Havre, 76058  Le
Havre Cedex, France
}
\begin{document}
\maketitle
\begin{abstract}
The automata arising from the well known conversion of regular expression to non deterministic automata have rather 
particular transition graphs. We refer to them as the Glushkov graphs, to honour his nice expression-to-automaton 
algorithmic short cut \cite{Glu61}. The Glushkov graphs have been characterized \cite{CZ97} in terms of simple graph 
theoretical properties and certain reduction rules. We show how to carry, under certain restrictions, this 
characterization over to the weighted Glushkov graphs. With the weights in a semiring $\K$, they are defined as the transition 
Glushkov $\K$-graphs of the Weighted Finite Automata (WFA) obtained by the generalized Glushkov construction \cite{CF03}  from the $\K$-expressions. It works provided that the semiring $\K$ is factorial and the $\K$-expressions are in the
 so called star normal form (SNF) of Br\"uggeman-Klein \cite{Bru93}. The restriction to the factorial semiring ensures to obtain algorithms. The restriction to the SNF would not be necessary 
if every $\K$-expressions were equivalent to some with the same litteral length, as it is the case for the 
boolean semiring $\B$ but remains an open question for a general $\K$.  
\end{abstract}

\section{Introduction}
The extension of boolean algorithms (over languages) to multiplicities
(over series)
has always been a central point in theoretical research. First,
Sch\"utzenberger \cite{Sch61.1} has given an equivalence between rational
and
recognizable series extending the classical result of Kleene \cite{Kle56}.
Recent contributions have been done in this area, an overview of knowledge of these domains is presented by Sakarovitch 
in \cite{Sak03}.
 Many research works have focused on producing a small WFA. For example, Caron and Flouret have extended the Glushkov construction to WFAs \cite{CF03}.
Champarnaud {\it et al} have designed a quadratic algorithm \cite{COZ09} for computing the equation WFA of a $\K$-expression. This equation WFA has been introduced by Lombardy and Sakarovitch as an extension of Antimirov's algorithm  \cite{LS01} based on partial derivatives.

Moreover, the Glushkov WFA of a $\K$-expression with $n$ occurrences of symbol
(we say that its alphabetic width is equal to $n$) 
has only $n+1$ states;
the equation $\K$-automaton
(that is a quotient of the Glushkov automaton)
has at most $n+1$ states.

On the opposite, classical algorithms compute $\K$-expressions
the size of which is exponential with respect to the number of states of the WFA. For example, let us cite the block decomposition algorithm  proven in \cite{BR88}.


 In this paper, we also address the problem of computing short $\K$-expressions,
and we focus on a specific kind of conversion based on Glushkov automata.
Actually the particularity of Glushkov automata is the fol\-lo\-wing:
any regular expression of width $n$ can be turned into its Glushkov $(n+1)$-state automaton;
if a $(n+1)$-state automaton is a Glushkov one, then it can be turned into an expression of width $n$.
The latter property is based on the characterization of the family of Glushkov automata in terms of graph properties presented in~\cite{CZ97}. These properties are stability, transversality and reducibility.
 Br\"uggemann-Klein defines regular expressions in {\it Star Normal Form} (SNF) \cite{Bru93}. These expressions are characterized by underlying Glushkov  automata where each edge is generated exactly one time. This definition is extended to multiplicities. The study  of the SNF case would not be necessary 
if all $\K$-expressions were equivalent to some in SNF with the same litteral length, as it is the case for the boolean semiring $\B$.
 
 The aim of this paper is to extend the characterization of Glushkov automata to the multiplicity case in order to compute a $\K$-expression of width $n$ from a $(n+1)$-state WFA. This extension requires to restrict the work to factorial semirings as well as Star Normal Form $\K$-expressions. 
 
We exhibit a procedure that, given a WFA $M$ on $\K$ a factorial semiring, outputs the following: 
either $M$ is obtained by the Glushkov algorithm from a proper $\K$-expression $E$ in Star Normal Form and the 
procedure computes a $\K$-expression $F$ equivalent to $E$, or $M$ is not obtained in that way and the procedure 
says no.

The following section recalls fundamental notions concerning automata, expressions and Glushkov conversion for both boolean and multiplicity cases. An error in the paper by Caron and Ziadi \cite{CZ97} is pointed out and corrected. The section 3 is devoted to the reduction rules for acyclic $\K$-graphs. Their efficiency is provided by the confluence of $\K$-rules. The next section gives orbit properties for Glushkov $\K$-graphs. The section 5 presents the algorithms computing a $\K$-expression from a Glushkov $\K$-graph and details an example.



\section{Definitions}
\subsection{Classical notions}
Let $\Sigma$ be a finite set of letters (alphabet), $\varepsilon$ the empty word
and $\emptyset$ the empty set.
Let ($\K$, $\oplus$, $\otimes$) be a  zero-divisor free semiring where
$\overline{0}$ is the neutral element of
$(\K,\oplus)$ and $\overline{1}$ the one of $(\K,\otimes)$.  The semiring $\K$
is said to be zero-divisor free \cite{HW96} if $\b{0}\neq \b{1}$ and if $\forall
x,y\in \K$, $x\otimes y=\b{0} \Rightarrow x=\b{0}\mbox{ or } y=\b{0}$.

A {\it formal series} \cite{BR88} is a mapping $S$
from $\Sigma^*$ into $\K$ usually denoted by $\displaystyle S=\sum_{w
\in \Sigma^*}
S(w)  w$ where $  S(w) \in \K$ is the
coefficient of $w$ in $S$. The {\it support} of $S$ is the language $Supp(S)=\{w
\in \Sigma^* | S(w) \neq \b{0}\}$.

In \cite{LS01}, Lombardy and Sakarovitch explain in details the computation of
$\K$- expressions. We have followed their model of grammar. Our constant symbols
are $\varepsilon$ the empty word and $\emptyset$. Binary rational operations are still $+$ and
$\cdotp$, the unary ones are Kleene closure $*$, positive closure $^+$ and for
every $k\in \K$, the multiplication to the left or to the right of an
expression $\times$. For an easier reading, we will write $kE$ (respectively $Ek$) for $k\times E$ (respectively 
$E\times k$). Notice that our definition of $\K$-expressions, which set is denoted
$E_{\K}$, introduces the operator of positive closure. This operator preserves
rationality with the same conditions (see below) that the Kleene closure's one.

$\K$-expressions are then given by the following grammar:
$$E\rightarrow a\in \Sigma \ |\ \emptyset \ |\ \varepsilon \ |\ (E+E)\ |\
(E\cdotp E)\ |\ (E^*)\ |\ (E^+)\ |\ (kE), k\in\K \ |\ (Ek) ,k\in K$$

Notice that parenthesis will be omitted when not necessary. The expressions $E^+$ and $E^*$ are called {\it closure expressions}.
If a series $S$ is represented by a $\K$-expression $E$, then we denote by $c(S)$ (or $c(E)$) the coefficient of the empty word of $S$.
A $\K$-expression $E$ is {\it valid} \cite{Sak03} if for each closure subexpression $F^*$ and $F^+$ of $E$, $\displaystyle \sum _{i=0}^{+\infty}c(F)\in \K$.
 
 A $\K$-expression $E$ is {\it proper} if for each closure subexpression $F^*$ and $F^+$ of $E$, $c(F)=\b{0}$.

We denote by ${\cal E}_\K$ the set of proper $\K$-expressions. Rational series
can then be defined as formal series expressed by {\it proper $\K$-expressions}.
For $E$ in ${\cal E}_\K$, $Supp(E)$ is the support of the rational series
defined by $E$.

The length of a $\K$-expression $E$, denoted by $||E||$, is the number of occurences of letters and of $\epsilon$ 
appearing in $E$. By opposition, the litteral length, denoted by $|E|$ is the number of occurences of letters in $E$.
 For example, the expression  $E=(a+3)(b+2)+ (-1)$ as a length of $5$ and a litteral length of $2$.

 A {\it weighted finite automaton} ({\it WFA}) on a zero-divisor free semiring
$\K$  over an alphabet $\Sigma$ \cite{Eil74} is a
$5$-tuple $(\Sigma ,Q,I,F,\delta)$  where $Q$ is a finite set of states and the
sets $I$,
$F$ and $\delta$ are mappings $I: Q \rightarrow \K $ (input weights),
$F: Q\rightarrow  \K$ (output weights), and $\delta : Q \times \Sigma \times
Q\rightarrow \K$ (transition weights).
The set of WFAs on $\K$ is denoted by $\M_{\K}$.
A WFA is {\it homogeneous} if all vertices reaching a same state are labeled by the same letter. 

A {\it $\K$-graph} is a graph $G=(X,U)$ labeled with coefficients in $\K$ where $X$ is the set of vertices and $U : X\times X \rightarrow \K$ is the 
function that associates each edge with its label in $\K$. When there is no edge from $p$ to $q$, we have $U(p,q)=\overline{0}$.
In case $\K=\B$, the boolean semiring,  $E_\B$ is the set of {\it regular
expressions} and, as the only element of $\K\setminus \b{0}$ is $\b{1}$, we omit
the use of coefficient and of the external product ($\b{1}a= a\b{1}=a$).
For a rational series $S$ represented by  $E\in  E_\B$, $Supp(E)$ is usually called the
language of $E$, denoted by $L(E)$ and $S=Supp(S)=L(E)$.
A boolean automaton (automaton in the sequel) $M$ over
an alphabet $\Sigma$ is usually defined
\cite{Eil74,HU79} as a $5$-tuple $(\Sigma ,Q,I,F,\delta)$ where $Q$ is a
finite set of states, $I \subseteq Q$ the set of initial states, $F
\subseteq Q$ the set of final states, and $\delta \subseteq Q \times
\Sigma \times Q $ the set of edges. We denote by  $L(M)$ the language
recognized by the automaton $M$. A graph $G=(X,U)$ is a $\B$-graph for which labels of edges are not written.

\subsection{Extended Glushkov construction}

An algorithm given by Glushkov \cite{Glu61} for computing an automaton with $n+1$ states from a regular expression of litteral length  $n$ has been extended to semirings $\K$ by the authors \cite{CF03}. Informally, the principle is to associate exactly one state in the computed automaton to each occurrence of letters in the expression. Then, we link by a transition two states of the automaton if the two  occurences of the corresponding letters in the expression can be read successively.

In order to recall  the extended Glushkov construction, we have to first define the ordered pairs and the supported operations.
An ordered pair $(l,i)$ consists of a coefficient $l\in \K\setminus \{\b{0}\}$ and a position $i\in \N$.
We also define the functions ${\cal I}_H:H\rightarrow \K$ such that ${\cal I}_H(i)$ is equal to $\b{1}$ 
if $i\in H$ and $\b{0}$ otherwise.
We define $P:2^{\K\setminus \{\b{0}\}\times \N}\rightarrow 2^\N$ the function that extracts positions from a set of ordered pairs  as follows: for $Y$ a set of ordered pairs, $P(Y)=\{i_j, 1\leq j\leq |Y|\mid \exists (l_j,i_j)\in Y\}$.

The function $\mbox{{\it Coeff}}_Y:P(Y) \rightarrow \K\setminus \{\b{0}\}$  extracts the coefficient associated to a position $i$ as follows: $\mbox{{\it Coeff}}_Y(i)=l$ for $(l,i)\in Y$.

Let  $Y,Z\subset \K\setminus \{\b{0}\}\times \N$ be two sets of ordered pairs.
We define the product of $k\in \K \setminus \b{0}$ and $Y$ by  $k\cdotp Y=\{(k\otimes l,i)\mid (l,i)\in Y\}$ and 
$Y\cdotp k=\{(l\otimes k,i) \mid (l,i)\in Y\}$, $\b{0}\cdotp Y=Y\cdotp \b{0}=\emptyset$. 
 We define the operation $\uplus$ by $Y\uplus Z = \{(l,i)\mid \mbox{either }(l,i)\in Y\mbox{ and }i\not\in P(Z)\mbox{ or }(l,i)\in 
Z\mbox{ and }i\not\in P(Y)\mbox{ or }(l_s,i)\in Y, (l_t,i)\in Z$ for some $l_s,l_t\in \K\mbox{ with }l=l_s\oplus l_t\neq
 \b{0}\}$. 

As in the original Glushkov construction \cite{Glu60,MY60}, and in order to specify their position in the expression, 
letters are subscripted following the order of reading. The resulting expression is denoted  $\b{E}$, defined over the 
alphabet of indexed symbols $\b{\Sigma}$, each one appearing at most once in $\b{E}$. The set of indices thus obtained is 
called positions and denoted by $\P{E}$. For example, starting from 
$E=(2a+b)^*\cdotp a\cdotp 3 b$, one obtains the indexed expression $\b{E}=(2a_1+b_2)^*\cdotp a_3\cdotp 3 b_4$, 
$\b{\Sigma}=\{a_1,b_2,a_3,b_4\}$ and  $\P{E}=\{1,2,3,4\}$.
Four functions are defined in order to
compute a WFA which needs not be deterministic. $\F{E}$
represents the set of initial positions of words of $Supp(\b{E})$ associated with their input weight,
$\L{E}$ represents the set of final positions of words of $Supp(\b{E})$ associated to their output weight and
$\Fol{E}{i}$ is the set of positions of words of $Supp(\b{E})$ which immediately follows
position $i$ in the expression $\b{E}$, associated to their transition weight.  In the boolean case, these sets are subsets of $\P{E}$. The $\Nl{E}$ set represents 
the coefficient of the empty word. The way to compute these sets is completely formalized in table \ref{tableset}. 

\begin{table}[H]
{\small
\begin{tabular}{|c||c|c|c|c|}
\hline
\bf{E}&\bf{Null(E)}&\bf{First(E)}&\bf{Last(E)}&\bf{Follow}(E,i)\\
\hline
\hline
\bf{$\emptyset$} &$\b{0}$&$\emptyset$&$\emptyset$&$\emptyset$\\
\hline
\bf{$\varepsilon$}&$\b{1}$&$\emptyset$&$\emptyset$&$\emptyset$\\
\hline
\bf{$a_j$}&$\b{0}$&$\{(\b{1},j)\}$&$\{(\b{1},j)\}$&$\emptyset$\\
\hline
\bf{$k F$}&$k\otimes \Nl{F}$&$k\cdotp \F{F}$&$\L{F}$&$\Fl{F}$\\
\hline
\bf{$F k$}&$\Nl{F}\otimes k$&$\F{F}$&$\L{F}\cdotp k$&$\Fl{F}$\\
\hline
\bf{$F+G$}&\begin{minipage}{1.8cm}\begin{center}$\Nl{F}$\\$\oplus$\\$ \Nl{G}$\end{center}\end{minipage}&\begin{minipage}
{2.7cm}\begin{center}$\F{F}$\\$\uplus$\\$\F{G}$\end{center}\end{minipage}&\begin{minipage}{2.5cm}\begin{center}$\L{F}$\\$
\uplus$\\$\L{G}$\end{center}\end{minipage}&\begin{minipage}{3.7cm}\begin{center}$\Y{\P{F}}\cdotp\Fl{F}$\\$\uplus$\\$\Y{\P{G}}
\cdotp\Fl{G}$\end{center}\end{minipage}\\
\hline
\bf{$F\cdotp G$}&\begin{minipage}{1.8cm}\begin{center}$\Nl{F}$\\$\otimes$\\$ \Nl{G}$\end{center}\end{minipage}
&\begin{minipage}{2.7cm}\begin{center}$\F{F}$ \\ $\uplus$\\ $\Nl{F}\cdotp\F{G}$\end{center}\end{minipage}&\begin{minipage}
{2.5cm}\begin{center}$\L{F}\cdotp \Nl{G}$\\ $\uplus$ \\ $\L{G}$\end{center}\end{minipage}&\begin{minipage}{3.7cm}
\begin{center}$\Y{\P{F}}\cdotp\Fl{F}$\\ $\uplus$ \\ $\Y{\P{G}}\cdotp\Fl{G}$\\ $\uplus$\\ $ \mbox{Coeff}_{\L{F}}(i)
\cdotp\F{G}$\end{center}\end{minipage}\\
\hline
\bf{$F^+$}&$\b{0}$&$\F{F}$&$\L{F}$&\begin{minipage}{3.7cm}\begin{center}$\Fl{F}$\\$\uplus$\\$ \mbox{Coeff}_{\L{F}}(i)
\cdotp\F{F}$\end{center}\end{minipage}\\
\hline
\bf{$F^*$}&$\b{1}$&$\F{F}$&$\L{F}$&\begin{minipage}{3.7cm}\begin{center}$\Fl{F}$\\$\uplus$\\$ \mbox{Coeff}_{\L{F}}(i)
\cdotp\F{F}$\end{center}\end{minipage}\\
\hline
\end{tabular}
}
\caption{Extended Glushkov functions}\label{tableset}
\end{table}

These functions allow us to define the WFA
$\b{M}=(\b{\Sigma},Q,\{s_I\},F,\b{\delta})$ where
\begin{enumerate}
\item $\b{\Sigma}$ is the indexed alphabet,
\item $s_I$ is the single initial state with no incoming edge with $\b{1}$ as input weight,
\item $Q=\P{E}\cup \{s_I\}$
\item $F:Q\rightarrow \K$ such that
$F(i)=\left\{
\begin{array}{ll}
\Nl{E}&\mbox{ if }i=s_I\\
\mbox{{\it Coeff}}_{\L{E}}(i)&\mbox{ otherwise}\\
\end{array}
\right.$
\item $\delta :Q\times \b{\Sigma}\times Q\rightarrow \K$ such that  $\delta(i,a_j,h)=\b{0}$ for every $h\neq j$, whereas\\
$\delta(i,a_j,j)=\left\{\begin{array}{cc}
\mbox{{\it Coeff}}_{\F{E}}(j) & i= s_I\\
\mbox{{\it Coeff}}_{\Fol{E}{i}}(j) & i\neq s_I \\
\end{array}
\right.$
\end{enumerate}

The Glushkov WFA $M=(\Sigma,Q,\{s_I\},F,\delta)$ of $E$ is computed from $\b{M}$ by
replacing the indexed letters on edges by the corresponding letters in
the expression $E$. We will denote $A_\K: {\cal E}_\K\rightarrow \M_\K$ the application such that $A_\K(E)$ is the Glushkov WFA obtained from $E$ by this algorithm proved in \cite{CF03}.


In order to compute a $\K$-graph from an homogeneous WFA $M$, we have to add a new vertex $\{\Phi \}$. Then $U$, the set of edges, is obtained from transitions of $M$ by removing labels and adding directed edges from every final state to $\{\Phi \}$. We label  edges to $\Phi$ with output weights of final states. The labels of the edges  $U(i,p)$ for $i\in Q$, $I(i)\neq \b{0}$, $p\in Q\cup \{\Phi\}$ are $\otimes$-multiplied by the input value of the initial state $i$ of $M$.

In case $M$ is a Glushkov WFA of a $\K$-expression $E$, the $\K$-graph obtained from $M$ is called  Glushkov $\K$-graph of $E$ and is denoted by $G_\K(E)$.

\subsection{Normal forms and casting operation}
\subsubsection*{Star normal form and epsilon normal form}
For the boolean case, Br\"uggemann-Klein defines regular expressions in {\it Star Normal Form} (SNF) \cite{Bru93} as expressions $E$ for which, for each
 position $i$ of $\P{E}$, when computing the $\Fol{E}{i}$ function, the unions of sets are disjoint. This definition is given only for usual 
operators ,$+$, $\cdotp $, $*$. We can extend this definition to the positive closure, $^+$ as follows:
\begin{definition}
A $\B$-expression $E$ is in SNF if, for each closure $\B$-subexpression $H^*$ or $H^+$, the SNF conditions 
(1) $\Fol{H}{\L{H}}\cap \F{H}=\emptyset$ and (2) $\varepsilon \not\in L(H)$ hold.
\end{definition} 
Then, the properties of the star normal form (defined with the positive closure) are preserved.

In the same paper, Br\"uggemann-Klein defines also the {\it epsilon normal form} for the boolean case. We extend this epsilon normal form to the 
positive closure operator.
\begin{definition}
The epsilon normal form for a $\B$-expression $E$ is defined by induction in the following way:
\begin{itemize}
\item{$[E=\varepsilon$ or $E=a]$} $E$ is in epsilon normal form.
\item{$[E=F+G]$} $E$ is in epsilon normal form if $F$ and $G$ are in epsilon normal form and if 
$\varepsilon\not\in L(F)\cap L(G)$.
\item{$[E=FG]$} $E$ is in epsilon normal form if $F$ and $G$ are in epsilon normal form.
\item{$[E=F^+$ or $E=F^*]$} $E$ is in epsilon normal form if $F$ is in epsilon normal form and $\varepsilon \not\in L(F)$.
\end{itemize}
\end{definition}

\begin{theorem}[\cite{Bru93}]
 For each regular expression $E$, there exists a regular expression $E^\bullet$ such that 
 \begin{enumerate}
 \item $A_{\B}(E)=A_{\B}(E^\bullet)$,
\item $ E^\bullet$ is in SNF
\item $E^\bullet$ can be computed from $E$ in linear time.
\end{enumerate}
\end{theorem}
Br\"uggemann-Klein has given every step for the computation of $E^\bullet$. This computation remains. We just have to add 
for $H^+$ the same rules as for $H^*$. Main steps of the proof are similar.\\

We extend the star normal form to multiplicities in this way. Let $E$ be a $\K$-expression. 
For every subexpression $H^*$ or $H^+$ in $E$, for each $x$ in $P(\L{H})$, 
$$P(\Fol{H}{x})\cap P(\F{H})=\emptyset$$
We do not have to consider the case of the empty word because $H^+$ and $H^*$ are proper $\K$-expressions if $c(H)= 0$.
  
As an example, let $\b{H}=2 a_1^++(3 b_2)^+$ and $\b{E}=(\b{H})^*$. We can see that the 
expression $\b{E}=(2 a_1^++(3 b_2)^+)^*$ is not in SNF, because
$2\in P(\L{H})$, $2\in $ {\it P(Follow(H,2))}$\cap P(\F{H})$. 

\subsubsection*{The casting operation $\sim$}
We have to define the casting $\sim$: $\M_{\K}\rightarrow \M_{\B}$.  This is similar to the way in which  Buchsbaum et al. \cite{BGW00} define the topology of a graph.
 A WFA $M=(\Sigma,Q,I,F,\delta)$ is casted into an automaton $\widetilde{M}=(\Sigma,Q,\widetilde{I},\widetilde{F},\widetilde{\delta})$ in the following way:
$\widetilde{I}, \widetilde{F}\subset Q$, $\widetilde{I}=\{q\in Q\mid
I(q)\neq \overline{0}\}$, $\widetilde{F}=\{q\in
Q\mid F(q)\neq \overline{0}\}$ and $\widetilde{\delta}=\{(p,a,q)\mid p,q\in Q,\ a\in
\Sigma
\mbox{ and } \delta((p,a,q)) \neq \overline{0}\}$.
 The casting operation can be extended to $\K$-expressions $\sim :{\cal E}_\K\rightarrow E_\B$.
The regular expression $\widetilde{E}$ is obtained from $E$ by replacing each $k\in \K \setminus \b{0}$ by $\b{1}$.
The $\sim$ operation on $E$ is an embedding of $\K$-expressions 
into regular ones. Nevertheless, the Glushkov $\B$-graph computed from a $\K$-expression $E$ may be different whether the Glushkov construction is applied first or the casting operation $\sim$. This is due to properties of $\K$-expressions. For example, let $\K=\Q$, $E=2a^*+(-2)b^*$ ($E$ is not in epsilon normal form). We then have $\widetilde{E}=a^*+b^*$. We can notice that $\widetilde{ A_{\K} }(E) \neq A_{\B}(\widetilde{E})$ ($E$ does not recognize $\epsilon$ but  $\widetilde{E}$ does).\\


\begin{lemma}\label{rat1} 
Let $E$ be a $\K$-expression.
 If $E$ is in SNF and in epsilon normal form, then $$\widetilde{ A_{\K} }(E) =A_{\B}(\widetilde{E}).$$
\end{lemma}

\noindent
{\bf Proof }We have to show that the automaton obtained by the Glushkov construction for an expression 
$E$ in ${\cal E}_\K$ has the same edges as the Glushkov automaton for $\widetilde{E}$.
First, we have $\P{E} = \P{\widetilde{E}}$, as $\widetilde{E}$ is obtained from $E$ only by deleting coefficients.
Let us show that $\F{\widetilde{E}}= \mbox{\it P}(\F{E})$ (states reached from the initial state)
by induction on the length of $E$. If $E = \varepsilon$, 
$\widetilde{E}=\epsilon$, $\F{\widetilde{E}} = \emptyset = \F{E} = \mbox{\it P} (\F{E})$. 
If $E = a \in \Sigma$, $\b{E}=a_1$ then $E = \widetilde{E}$, $\F{E} = \{(\b{1}, 1)\}$, $\mbox{\it P}(\F{E}) = \{1\} 
= \F{\widetilde{E}}$. 
Let $F$ satisfy the hypothesis, and $E= k F $,$k\in\K \setminus{\b{0}}$. In this case, $\widetilde{E}=\widetilde{F}$, 
$\mbox{\it P}(\F{E}) = \mbox{\it P}(k .\F{F}) = \mbox{\it P}(\F{F}) = \F{\widetilde{F}} = \F{\widetilde{E}}$.
 If $E = F k$, $k \in\K$, $\widetilde{E}=\widetilde{F}$, 
 $\mbox{\it P}(\F{E}) = \mbox{\it P}(\F{F}) = \F{\widetilde{F}} = \F{\widetilde{E}}$.
 
 If $E = F +H$, and if $F$ and $H$ satisfy the induction hypothesis, and as the coefficient of the empty word is $\b{0}$ for one of the two subexpression $F$ or $H$ (epsilon normal form), we have 
 $ \widetilde{E}=\widetilde{F} +\widetilde{H} $, $\F{\widetilde{F} +\widetilde{H}}=
 \F{\widetilde{F}} \cup \F{\widetilde{H}} = \mbox{\it P}(\F{F}) \cup \mbox{\it P}(\F{H})$
 which is equal to  $ \mbox{\it P}(\F{F+H})$ by induction. We obtain the same result concerning $F\cdot H$, $F^+$ and $F^*$.

\noindent 
The equality $\L{\widetilde{E}}= \mbox{\it P}(\L{E})$ is obtained similarly.
 
The last function used to compute the Glushkov automaton is the {\it Follow} function. 
Let $E$ be a $\K$-expression and  $i \in \P{E}$. If $E = \varepsilon$, $\widetilde{E}=\epsilon$, 
$\Fol{\widetilde{E}}{i} =\emptyset =\Fol{E}{i} = \mbox{\it P}(\Fol{E}{i})$. 
If $E = a \in \K$, $E = \widetilde{E}$, $\Fol{\widetilde{E}}{i} =\emptyset $.  
Let $F$ satisfy $\Fol{\widetilde{F}}{i}= \mbox{\it P}(\Fol{F}{i})$ for all $i \in \P{F}$.  
If $E$ is $k F$ or $F k$, $k \in\K \setminus{\b{0}}$, $ \mbox{\it P}(\Fol{E}{i}) = \mbox{\it P}(\Fol{F}{i}) = \Fol{\widetilde{F}}{i}$ by hypothesis. If $F$ and $H$ satisfy the induction  hypothesis, and if $E = F +H$, (and $i \in \P{F}$ without loss of generality), $\Fol{F+H}{i} =\Fol{F}{i}$, then $\mbox{\it P}(\Fl{F}) = \Fol{\widetilde{F}}{i} $. We obtain similar results for $E = F.H$ as there is no intersection between positions of $F$ and $H$.
Concerning the star operation, let $E = F^*$, with $\Fol{\widetilde{F}}{i}= \mbox{\it P}(\Fol{F}{i})$ for all $i \in \P{F}$. Then, $\mbox{\it P}(\Fol{F^*}{i})= \mbox{\it P}(\Fol{F}{i} \cup \mbox{Coeff}_{\L{F}}(i)\cdotp\F{F} )$.   But by definition, as $F$ is in SNF, we know that $\Fol{F}{i}\cap \F{F}=\emptyset$, so $\mbox{\it P}(\Fol{F^*}{i})= \Fol{\widetilde{F^*}}{i}$. In fact, it means that if there exists a couple $(\alpha,j)\in \Fol{F}{i}$, there cannot exist $(\beta,j)\in \F{F}$. Otherwise, the expression would not be in SNF, and it would be possible that $\beta = \alpha$, which would make $j \not\in \P{F^*}$ and imply a deletion of an edge. A same reasonning can be done for the positive closure operator.

Hence, the casting operation $\sim$ and the Glushkov construction commute for the composition operation if we do not consider the empty word.
\cqfd

\subsection{Characterization of Glushkov automata in the boolean case}\label{Glushkov-Charac}
The aim of the paper by Caron and Ziadi  \cite{CZ97} is to know how boolean Glushkov graphs can be characterized. We recall here the definitions which allow us to give the main theorem of their paper. These notions will be necessary to extend this characterization to Glushkov $\K$-graphs.

%
%
A {\it hammock} is a graph  $G=(X,U)$ without a loop if $|X|=1$, otherwise  it has two distinct  vertices  $i$ and $t$ such that, for any vertex  $x$ of $X$,
(1) there exists a path from  $i$ to  $t$ going through $x$,
(2) there is no non-trivial path from  $t$ to $x$ nor from  $x$ to $i$. Notice that every hammock with at least two vertices has a unique root (the vertex $i$) and anti-root (the vertex $t$). 

%
%
Let $G=(X,U)$ be a hammock.
 We define $\O=(X_\O,U_\O) \subseteq G$ as an {\it orbit} of $G$ if and only if  for all $x$ and $x'$ in $X_\O$ there exists a non-trivial path from $x$ to $x'$.
The orbit $\O$ is {\it maximal}  if, for each vertex $x\in X_\O$ and for each vertex $x'\in X\setminus X_\O$, there do not exist both a path from $x$ to $x'$ and a path from $x'$ to $x$. Equivalently, $\O \subseteq G$ is a maximal orbit of $G$ if and only if it is a strongly connected component with at least one edge.

Informally, in a Glushkov graph obtained from a regular expression $E$, the set of vertices of a maximal orbit corresponds exactly to the set of positions of a closure subexpression of $E$.

The set of direct successors (respectively direct predecessors) of $x\in X$ is denoted by $Q^+(x)$  (respectively $Q^-(x)$). Let $n_x=|Q^-(x)|$ and $m_x=|Q^+(x)|$. 
For an orbit  $\O\subset G$, $\O^+(x)$ denotes $Q^+(x)\cap (X\setminus \O)$ and $\O^-(x)$ denotes the set $Q^-(x)\cap (X\setminus \O)$.
In other words, $\O^+(x)$ is the set of vertices which are directly reached from $x$ and which are not in $\O$. By extension, $\O^+=\bigcup_{x\in \O} \O^+(x)$ and  $\O^-=\bigcup_{x\in \O} \O^-(x)$.
%
%
 The sets $\E{\O}=\{x\in X_\O \mid \O^-(x)\neq \emptyset\}$  and $\S{\O}=\{x\in X_\O \mid  \O^+(x)\neq \emptyset\}$
denote  the {\it input} and the {\it output} of the orbit $\O$. As $G$ is a hammock, $\E{\O}\neq \emptyset$ and  
$\S{\O}\neq \emptyset$. 
%
%
An orbit $\O$ is  {\it stable} if $\S{\O}\times \E{\O}\subset U$.
%
%
 An orbit $\O$ is  {\it transverse} if, for all $x,y \in \S{\O}$, $\O^+(x)=\O^+(y)$ 
and, for all $x,y \in \E{\O}$,  $\O^-(x)=\O^-(y)$. 

%
%
An orbit $\O$ is {\it strongly stable} (respectively {\it strongly transverse}) if it is stable (respectively transverse)  and if after deleting the edges  in $\S{\O}\times \E{\O}$ (1) there does not exist any suborbit $\O'\subset \O$ or (2) every maximal suborbit of $\O$  is strongly  stable (respectively strongly transverse).
%
%
The hammock $G$ is stronly stable (respectively strongly transverse) if (1) it has no orbit or (2) every maximal orbit $\O\subset G$ is strongly stable (respectively strongly transverse).

%
%
 If  $G$ is  strongly  stable, then we call {\it the graph without orbit} of $G$, denoted by $SO(G)$, the acyclic directed graph obtained by recursively deleting, for every maximal orbit  $\O$ of $G$, the edges in $\S{\O}\times \E{\O}$. The graph $SO(G)$ is then reducible if it can be reduced to one vertex by iterated applications of the three following rules:
\begin{itemize}
\item {\bf Rule {\bf $R_1$}}: If $x$ and $y$ are vertices such that $Q^-(y)=\{x\}$ and $Q^+(x)=\{y\}$, then delete $y$ and define $Q^+(x):=Q^+(y)$.
\item{\bf Rule  {\bf $R_2$}}: If $x$ and $y$ are vertices such that $Q^-(x)=Q^-(y)$ and $Q^+(x)=Q^+(y)$, then delete $y$ and any edge connected to $y$.	
\item{\bf Rule {\bf $R_3$}}: If $x$ is a vertex such that for all $y\in Q^- (x),\; Q^+(x)\subset Q^+(y)$, then delete edges in $Q^-(x) \times Q^+(x)$.  
\end{itemize}
 
\begin{theorem}[\cite{CZ97}]\label{th}
$G=(X,U)$ is a Glushkov graph if and only if the three following conditions are satisfied: 
\begin{itemize}
\item $G$ is a  hammock.
\item Each maximal orbit in G is  {\it strongly  stable} and {\it strongly transverse}.
\item The graph without orbit $SO(G)$ is {\it reducible}.
\end{itemize}
\end{theorem}

\subsection{The problem of reduction rules}
\subsubsection*{An erroneous statement in the paper by Caron and Ziadi}
In \cite{CZ97}, the definition of the $R_3$ rules is wrong in some cases. Indeed, if we consider the regular expression $E=(x_1 + \epsilon)(x_2 + \epsilon)+(x_3 + \epsilon)(x_4 + \epsilon)$, the graph obtained from the Glushkov algorithm is as follows

\VCDraw[.85]{
\begin{VCPicture}{(-6,-3)(12,3)}
\State[1]{(3,-1.5)}{1}
\State[2]{(6,-1.5)}{2}
\State[3]{(3,1.5)}{3}
\State[4]{(6,1.5)}{4}
\State[s_I]{(0,0)}{si}
\State[\Phi]{(9,0)}{phi}

\EdgeR{si}{1}{}
\EdgeR{si}{2}{}
\EdgeR{si}{3}{}
\EdgeR{si}{4}{}
\EdgeR{si}{phi}{}
\EdgeR{1}{2}{}
\EdgeR{1}{phi}{}
\EdgeR{2}{phi}{}
\EdgeR{3}{4}{}
\EdgeR{3}{phi}{}
\EdgeR{4}{phi}{}
\end{VCPicture}
}

Let us now try to reduce this graph with the reduction rules as they are defined in \cite{CZ97}. We can see that the sequel of applicable rules is $R_3$, $R_3$ and $R_1$. We can notice that there is a multiple choice for the application of the first $R_3$ rule, but after having chosen the vertex on which we will apply this first rule, the sequel of rules leads to a single graph (exept with the numerotation of vertices).
\begin{figure}[H]
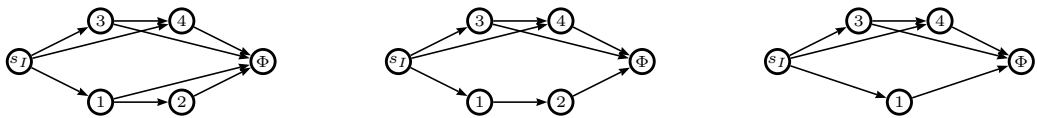

\VCDraw[.60]{
\begin{VCPicture}{(0,-3)(12,3)}
\SmallState
\tiny
\State[1]{(3,-1.5)}{1}
\State[2]{(6,-1.5)}{2}
\State[3]{(3,1.5)}{3}
\State[4]{(6,1.5)}{4}
\State[s_I]{(0,0)}{si}
\State[\Phi]{(9,0)}{phi}

\EdgeR{si}{1}{}
\EdgeR{si}{3}{}
\EdgeR{si}{4}{}
\EdgeR{1}{2}{}
\EdgeR{1}{phi}{}
\EdgeR{2}{phi}{}
\EdgeR{3}{4}{}
\EdgeR{3}{phi}{}
\EdgeR{4}{phi}{}

\State[1]{(17,-1.5)}{11}
\State[2]{(20,-1.5)}{12}
\State[3]{(17,1.5)}{13}
\State[4]{(20,1.5)}{14}
\State[s_I]{(14,0)}{1si}
\State[\Phi]{(23,0)}{1phi}

\EdgeR{1si}{11}{}
\EdgeR{1si}{13}{}
\EdgeR{1si}{14}{}
\EdgeR{11}{12}{}
\EdgeR{12}{1phi}{}
\EdgeR{13}{14}{}
\EdgeR{13}{1phi}{}
\EdgeR{14}{1phi}{}

\State[1]{(32.5,-1.5)}{21}

\State[3]{(31,1.5)}{23}
\State[4]{(34,1.5)}{24}
\State[s_I]{(28,0)}{2si}
\State[\Phi]{(37,0)}{2phi}

\EdgeR{2si}{21}{}
\EdgeR{2si}{23}{}
\EdgeR{2si}{24}{}
\EdgeR{21}{2phi}{}
\EdgeR{23}{24}{}
\EdgeR{23}{2phi}{}
\EdgeR{24}{2phi}{}
\end{VCPicture}
}
\caption{Application of $R_3$ on $1$, $R_3$ on $2$ and $R_1$ on $1$ and $2$.}
\end{figure}

We can see that the graph obtained is no more reducible. This problem is a consequence of the multiple computation of the edge $(0,\Phi)$. In fact, this problem is solved when each edge of the acyclic Glushkov graph is computed only once. It is the case when $E$ is in epsilon normal form.

\subsubsection*{A new $R_3$ rule for the boolean case}
Let $G=(X,U)$ be an acyclic graph. The rule  $R_3$ is as follows:\\
\begin{itemize}
\item If $x\in X$ is a vertex such that for all $y\in Q^- (x),\; Q^+(x)\subset Q^+(y)$, then delete the edge $(q^-,q^+)\in Q^-(x) \times Q^+(x)$ if there does not exist a vertex $z\in X\setminus \{x\}$ such that the following conditions are true:
\begin{itemize}
\item there is neither a path from $x$ to $z$ nor a path  from   $z$ to $x$,
\item $q^-\in Q^-(z)$ and $q^+\in Q^+(z)$,
\item $|Q^-(z)|\times |Q^+(z)|\neq 1$.
\end{itemize}
\end{itemize}
The new rule $R_3$ check whether conditions of the old $R_3$ rules are verified and moreover deletes an edge only if it does not correspond to the $\varepsilon$ of more than one subexpression.
The validity of this rule is shown in Proposition \ref{K-red}.

\section{Acyclic Glushkov WFA properties}
The definitions of section \ref{Glushkov-Charac} related to graphs are extended to $\K$-graphs by considering that edges labeled $\b{0}$ do not exist.

Let us consider $M$ a WFA without orbit.
Our aim here is to give conditions on weights in order to check whether $M$ is a Glushkov WFA. 
Relying on the boolean characterization, we can deduce that $M$ is homogeneous and that the Glushkov graph of $\widetilde{M}$ is reducible.

\subsection{$\K$-rules}

$\K$-rules can be seen as an extension of reduction rules. Each rule is divided into two parts: a graphic condition on edges, and a numerical condition (exept for the $\K R_1$-rule) on coefficients.
The following definitions allow us to give numerical constraints for the application of $\K$-rules.

Let $G=(X,U)$ be a $\K$-graph and let $x,y\in X$. Let us now define the set of beginnings of the set $Q^-(x)$ as $B(Q^-(x))\subseteq Q^-(x)$. A vertex  $x^-$ is in $B(Q^-(x))$ if  for all $q^-$ in $Q^-(x)$ there is not a non trivial path from $q^-$ to $x^-$. In the same way, we define the set of terminations of $Q^+(x)$ as  $T(Q^+(x))\subseteq Q^+(x)$.  A vertex  $x^+$ is in $T(Q^+(x))$ if for all $q^+$ in $Q^+(x)$ there is not a non trivial path from $x^+$ to $q^+$.

We say that $x$ and $y$ are {\it backward equivalent} if $Q^-(x)=Q^-(y)$ and there exist $l_x,l_y\in \K$ such that for every $q^-\in Q^-(x)$, there exists $\alpha_{q^-} \in \K$ such that $U(q^-,x)=\alpha_{q^-} \otimes l_x$ and $U(q^-,y)=\alpha_{q^-} \otimes l_y$. Similarly, we say that $x$ and $y$ are {\it forward equivalent} if $Q^+(x)=Q^+(y)$ and there exist $r_x,r_y\in \K$ such that for every $q^+\in Q^+(x)$, there exists $\beta_{q^+} \in \K$ such that $U(x,q^+)=r_x \otimes \beta _{q^+}$ and $U(y,q^+)=r_y \otimes \beta _{q^+}$. Moreover, if $x$ and $y$ are both backward and forward equivalent, then we say that $x$ and $y$ are {\it bidirectionally equivalent}.

 In the same way, we say that $x$ is {\it $\epsilon$-equivalent}  if for all $(q^-,q^+)\in Q^-(x)\times Q^+(x)$ the edge $(q^-,q^+)$ exists and if there exist $k,l,r\in \K$ such that for every $q^-\in Q^-(x)$ there exists $\alpha_{q^-}\in \K$ and for every $q^+\in Q^+(x)$ there exist $\beta_{q^+}\in \K$,  such that $U(q^-,x)=\alpha _{q^-}\otimes l$, $U(x,q^+)=r\otimes \beta _{q^+}$ and  $U(q^-,q^+)=\alpha _{q^-}\otimes k\otimes \beta _{q^+}$.
 
Similarly, $x$ is {\it quasi-$\epsilon$-equivalent} if 
\begin{itemize}
\item $B(Q^-(x))\neq Q^-(x)$  or $T(Q^+(x))\neq Q^+(x)$, and
\item for all $(q^-,q^+)\in Q^-(x)\times Q^+(x)\setminus B(Q^-(x))\times T(Q^+(x))$, the edge $(q^-,q^+)$ exists, and 
\item  there exist $k,l,r\in \K$ such that for every $q^-\in Q^-(x)$ there exist $\alpha _{q^-}\in \K$ and for every $q^+\in Q^+(x)$, there exist $\beta _{q^+} \in \K$ such that  $U(q^-,x)=\alpha _{q^-}\otimes l$, $U(x,q^+)=r\otimes \beta _{q^+}$, and
\item  if $q^-\not\in B(Q^-(x))$ or $q^+\not\in T(Q^+(x))$ 
\begin{itemize}
\item   then $U(q^-,q^+)=\alpha _{q^-}\otimes k\otimes \beta_{q^+}$ 
\item else  there exists $\gamma  \in \K$ such that $U(q^-,q^+)=\gamma \oplus \alpha_{q^-} \otimes k \otimes \beta_{q^+}$ (Notice that if the edge from $q^-$ to $q^+$ does not exist in the automaton, then $U(q^-,q^+)=\b{0}$ and it is possible to have $\gamma \oplus \alpha_{q^-} \otimes k \otimes \beta_{q^+}=\b{0}$).
\end{itemize}
\end{itemize}
In order to clarify our purpose, we have distinguished the case where $(q^-,q^+)$ are superpositions of edges (quasi-$\epsilon$-equivalence  of $x$) to the case where they are not ($\epsilon$-equivalence  of $x$).\\

\noindent
{\bf Rule} $\boldsymbol {\K R_ 1}$: If $x$ and $y$ are vertices such that $Q^-(y)=\{x\}$ and $Q^+(x)=\{y\}$, then delete $y$ and define $Q^+(x)\leftarrow Q^+(y)$.\\

\begin{figure}[H]
\centerline{\includegraphics[width=90mm]{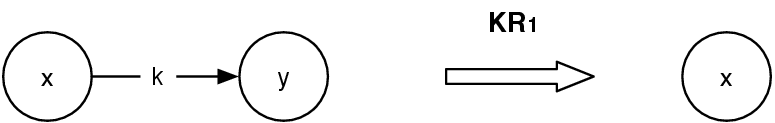}}
\caption{$\K R_1$ reduction rule}\label{KR1-rule}
\end{figure}
\noindent
{\bf Rule}  $\boldsymbol{\K R_2}$: If  $x$ and $y$ are bidirectionally equivalent, with $l_x, l_y, r_x, r_y \in \K$ are the constants satisfying such a definition,
then 
\begin{itemize}
\item delete $y$  and any edge connected to $y$ 
\item for every $q^-\in Q^-(x)$ and $q^+\in Q^+(x)$ set $U'(q^-,x)=\alpha _{q^-}$ and  $U'(x,q^+)=\beta  _{q^+}$ where $\alpha _{q^-}$ and $\beta _{q^+}$ are defined as in the bidirectional equivalence. 
\end{itemize}

\begin{figure}[H]
\includegraphics[width=140mm]{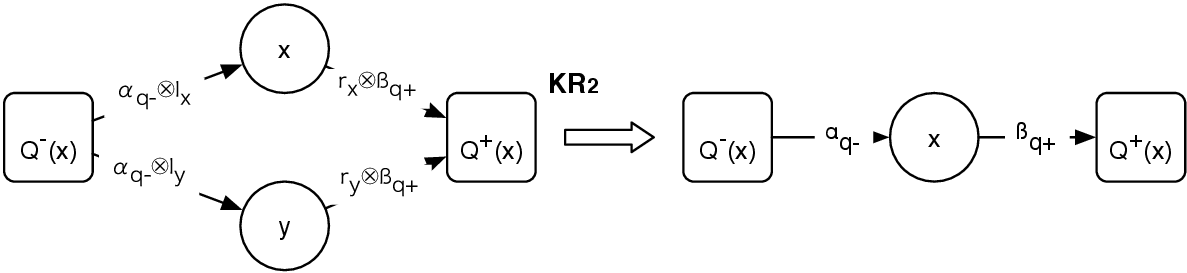}
\caption{$\K R_2$ reduction rule}\label{KR2-rule}
\end{figure}

\noindent
{\bf Rule} $\boldsymbol{\K R_3}$: If $x$ is $\epsilon$-equivalent or $x$ is quasi-$\epsilon$-equivalent with  $l, r, k, \gamma \in \K$ the constants satisfying such a definition,
then
\begin{itemize}
\item if $x$ is $\epsilon$-equivalent 
\begin{itemize}
\item then delete every $(q^-,q^+)\in Q^-(x)\times Q^+(x)$,
\item else delete every $(q^-,q^+)\in Q^-(x)\times Q^+(x)\setminus B(Q^-(x))\times T(Q^+(x))$.
\end{itemize}
\item for every $q^-\in Q^-(x)$ and $q^+\in Q^+(x)$ set $U'(q^-,x)=\alpha _{q^-}$ and  $U'(x,q^+)=\beta  _{q^+}$ where $\alpha _{q^-}$ and $\beta _{q^+}$ are defined as in the $\epsilon$-equivalence or quasi-$\epsilon$-equivalence. 
\item If $x$ is quasi-$\epsilon$-equivalent then compute the new edges from  $B(Q^-(x))\times T(Q^+(x))$ labeled $\gamma$.
\end{itemize}

\begin{figure}[H]
\includegraphics[width=140mm]{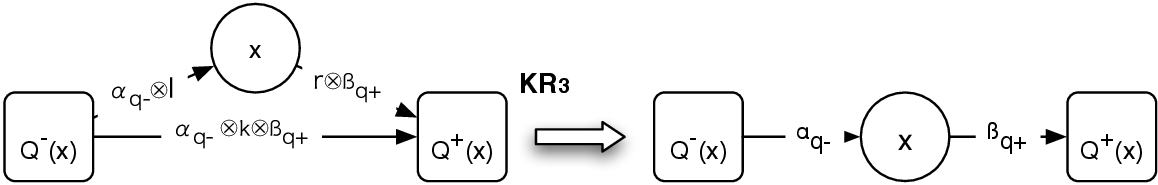}
\caption{$\K R_3$ reduction when $x$ is $\epsilon$-equivalent}\label{KR3a-rule}

\end{figure}

\begin{figure}[H]
\includegraphics[width=140mm]{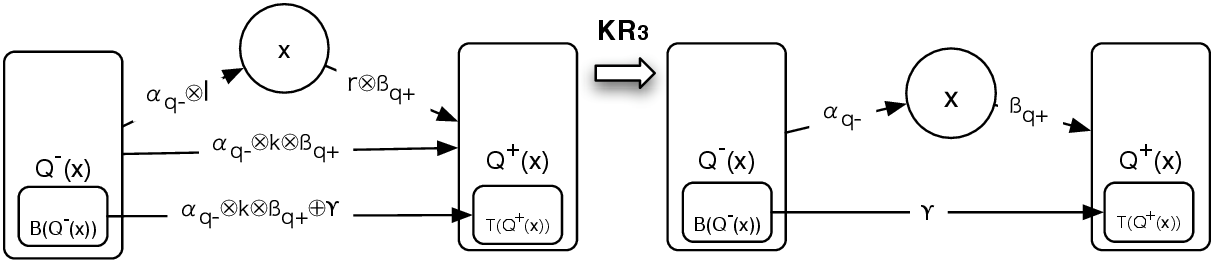}
\caption{$\K R_3$ reduction when $x$ is quasi-$\epsilon$-equivalent}\label{KR3b-rule}

\end{figure}

\subsection{Confluence for $\K$-rules}
In order to have an algorithm checking whether a $\K$-graph is a Glushkov $\K$-graph, we have to know (1) if it is decidable to apply a $\K$-rule on some vertices and (2) if the application of $\K$-rules ends. In order to ensure these characteristics, we will specify some sufficient properties on the semiring $\K$. Let us define $\K$ as a field or as a factorial semiring. A factorial semiring $\K$ is a zero-divisor free semiring for which every non-zero, non-unit element $x$ of $\K$ can be written as a product of irreducible elements of $\K$ $x=p_1\cdots p_n$, and this representation is unique apart from the order of the irreducible elements. This notion is a slight adaptation of the factorial ring notion.

 It is clear that, if $\K$ is a field, the application of $\K$-rules is decidable. Conditions of application of $\K$-rules are sufficient to define an algorithm. In the case of a factorial semiring, as the decomposition is unique, a $gcd$ is defined\footnote{In case $\K$ is not commutative, left $gcd$ and right $gcd$ are defined.} and it gives us a procedure  allowing us to apply one rule ($\K R_2$ or $\K R_3$) on a $\K$-graph if it is possible. It ensures the decidability of $\K$-rules application for factorial semirings.  For both cases (field and factorial semiring), we prove that $\K$-rules are confluent. It ensures the ending of the algorithm allowing us to know whether a $\K$-graph is a Glushkov one. \\


 We explicit algorithms in order to apply the $\K R_2$ and $\K R_3$ rules. Algorithm \ref{algoR2} tests whether the $\K R_2$-rule graphical and numerical conditions for two states are verified. If so, it returns the partially reduced $\K$-graph. 
 Algorithm \ref{algoR3} is divided into three functions. The first one check whether the $\K R_3$-graphical conditions are checked on a state $x$ ({\sc $\K R_3$GraphicalEquivalenceConditionsChecking}) and returns the $\epsilon$ or quasi-$\epsilon$-equivalence type of $x$. Then, depending on the type of $x$, the numerical conditions for $\epsilon$ or quasi-$\epsilon$-equivalence are verified (function {\sc EquivalenceChecking}). Finally a partially reduced $\K$-graph is obtained using {\sc GraphComputing} function.
 
{\small

\begin{algorithm}[H]
 \begin{algo}{$\K R_3$-Application}{x,G}
\IN { One state $x$ of a $\K$-graph $G=(X,U)$}
\OUT {The newly computed graph $G$}
\IF{\CALL{$\K R_3$GraphicalEquivalenceConditionsChecking}{x,G, type}=False}
\RETURN{False}
\FI
\COM{If type is equal to $\epsilon$ (resp. quasi-$\epsilon$) lines labeled \{{\bf quasi-$\epsilon$} \} (resp. \{ {\bf $\epsilon$} \})\CUT of the functions below are deleted}
\IF{\CALL{EquivalenceChecking}{x,G,[\alpha],[\beta],k,[\gamma]}=False}
\RETURN{False}
\FI
\CALL{GraphComputing}{x,G,[\alpha],[\beta],k,[\gamma]}
\RETURN{True}
\end{algo}
\caption{Application of the $\K R_3$ rule for a state}\label{algoR3}
\end{algorithm}
\begin{algorithm}[H]
 \begin{algo}{$\K R_2$-Application}{x,y,G}
\IN { Two states $x$ and $y$ of a $\K$-graph $G=(X,U)$}
\OUT {The newly computed graph $G$}
\IF{Q^-(x)\neq Q^-(y)\mbox{ or }Q^+(x)\neq Q^+(y)}
\RETURN{False}
\FI
\SET{q^-_1}{\mbox{a vertex of }Q^-(x)}
\SET{\mbox{gcd}_r(x)}{U(q_1^-,x)}
\SET{\mbox{gcd}_r(y)}{U(q_1^-,y)}
\DOFOREACH {q^-\in Q^-(x)}
\SET{\mbox{gcd}_r(x)}{\CALL{right gcd}{U(q^-,x),\mbox{gcd}_r(x)}}
\SET{\mbox{gcd}_r(y)}{\CALL{right gcd}{U(q^-,y),\mbox{gcd}_r(y)}}
\OD
\DOFOREACH {q^-\in Q^-(x)}
\STATE{compute $\alpha_{q^-}$ such that $U(q^-,x)=\alpha_{q^-}\otimes \mbox{gcd}_r(x)$}
\IF {\alpha_{q^-}\otimes \mbox{gcd}_r(y) \neq U(q^-,y)}
\RETURN{False}
\FI
\OD

\SET{q^+_1}{\mbox{a vertex of }Q^+(x)}
\SET{\mbox{gcd}_l(x)}{U(x,q_1^+)}
\SET{\mbox{gcd}_l(y)}{U(y,q_1^+)}
\DOFOREACH {q^+\in Q^+(x)}
\SET{\mbox{gcd}_l(x)}{\CALL{left gcd}{U(x,q^+),\mbox{gcd}_l(x)}}
\SET{\mbox{gcd}_l(y)}{\CALL{left gcd}{U(y,q^+),\mbox{gcd}_l(y)}}
\OD
\DOFOREACH {q^+\in Q^+(x)}
\STATE{compute $\beta_{q^+}$ such that $U(x,q^+)=\mbox{gcd}_l(x)\otimes \beta_{q^+}$}
\IF {\mbox{gcd}_l(y)\otimes \beta _{q^+}\neq U(y,q^+)}
\RETURN{False}
\FI
\OD

\STATE{delete $y$ and any edge connected to $y$}
\DOFOREACH {q^-\in Q^-(x)}
\SET{U(q^-,x)}{\alpha_{q^-}}
\OD
\DOFOREACH {q^+\in Q^+(x)}
\SET{U(q^+,x)}{\beta_{q^+}}
\OD
\RETURN{True}
\end{algo}
\caption{Application of the $\K R_2$ rule for two states}\label{algoR2}
\end{algorithm}

\begin{algo}{GraphComputing}{x,G,[\alpha],[\beta],k,[\gamma]}
\IN { One state $x$ of a $\K$-graph $G=(X,U)$ \CUT 
$\alpha\in \K^{|Q^-(x)|}$, $\beta\in \K^{|Q^+(x)|}$,$k\in \K$}
\RCOM{{\bf quasi-$\epsilon$}}
\IN { and $\gamma \in \K^{|B(Q^-(x))|\times |T(Q^+(x))|}$}
\OUT {The newly computed graph $G$}

\DOFOREACH{q^-\in Q^-(x)}
\SET {U(q^-,x)}{\alpha_{q^-}}
\OD
\DOFOREACH {q^+\in Q^+(x)}
\SET {U(x,q^+)}{\beta_{q^+}}
\OD
\RCOM{{\bf $\epsilon$}}
\STATE {delete any edge $(q^-,q^+)\in Q^-(x)\times Q^+(x)$}
\RCOM{{\bf quasi-$\epsilon$}}
\STATE {delete any edge $(q^-,q^+)\in Q^-(x)\times Q^+(x)\setminus B(Q^-(x))\times T(Q^+(x))$}
\RCOM{{\bf quasi-$\epsilon$}}
\DOFOREACH{(q^-,q^+)\in B(Q^-(x))\times T(Q^+(x))}
\RCOM{{\bf quasi-$\epsilon$}}
\SET {U(q^-,q^+)}{\gamma (q^-,q^+)}
\OD
\end{algo} 

\begin{algo}[noends]{$\K R_3$GraphicalEquivalenceConditionsChecking}{x,G, type}
\IN{ One state $x$ of a $\K$-graph $G=(X,U)$}
\OUT{type $\in$ $\{\epsilon$-equivalence, quasi-$\epsilon$-equivalence\}}
\STATE{compute $B(Q^-(x))$ and $T(Q^+(x))$}
\IF{ B(Q^-(x))=Q^-(x) \mbox{ and }T(Q^+(x))=Q^+(x)}
\DOFOREACH {q^-\in Q^-(x)}
\DOFOREACH {q^+\in Q^+(x)}
\IF{U(q^-,q^+)= \b{0}}
\RETURN{False}
\FI
\OD
\OD
\SET{\mbox{type}}{\epsilon}
\RETURN{True}
\ELSE
\DOFOREACH {q^-\in Q^-(x)}
\DOFOREACH {q^+\in Q^+(x)}
\IF{(q^-,q^+) \in Q^-(x) \times Q^+(x) \setminus B(Q^-(x))\times T(Q^+(x))
\CUT \mbox{and }U(q^-,q^+))= \b{0}}
\RETURN{False}
\FI
\OD
\OD
\SET{\mbox{type}}{\mbox{quasi-}\epsilon}
\RETURN{True}
\FI
\end{algo} 

\begin{algo}{EquivalenceChecking}{x,G,[\alpha],[\beta],k,[\gamma]}
\IN { One state $x$ of a $\K$-graph $G=(X,U)$}
\OUT{$\alpha\in \K^{|Q^-(x)|}$, $\beta\in \K^{|Q^+(x)|}$,$k\in \K$}
\RCOM{{\bf quasi-$\epsilon$}}
\OUT{ and $\gamma \in \K^{|B(Q^-(x))|\times |T(Q^+(x))|}$}
\SET{q^-_1}{\mbox{a vertex of }Q^-(x)}
\SET {\mbox{gcd}_r}{U(q_1^-,x)}
\DOFOREACH {q^-\in Q^-(x)}
\SET{\mbox{gcd}_r}{\CALL{right gcd}{U(q^-,x),\mbox{gcd}_r}}
\OD
\DOFOREACH {q^-\in Q^-(x)}
\STATE{compute $\alpha_{q^-}$ such that
$U(q^-,x)=\alpha_{q^-}\otimes \mbox{gcd}_r$}
\OD
\SET{q^+_1}{\mbox{a vertex of }Q^+(x)}
\SET {\mbox{gcd}_l}{U(x,q_1^+)}
\DOFOREACH {q^+\in Q^+(x)}
\SET{\mbox{gcd}_l}{\CALL{left gcd}{\mbox{gcd}_l,U(x,q^+)}}
\OD
\DOFOREACH {q^+\in Q^+(x)}
\STATE{compute $\beta_{q^+}$ such that
$U(x,q^+)=\mbox{gcd}_l\otimes \beta_{q^+} $}
\OD
\RCOM{{\bf $\epsilon$}}
\SET{(q_1^-,q_1^+)}{\mbox{ a couple of vertices of }Q^-(x)\times Q^+(x)}
\RCOM{{\bf quasi-$\epsilon$}}
\SET{(q_1^-,q_1^+)}{\mbox{ a couple of vertices of }\CUT Q^-(x)\times Q^+(x)\setminus B(Q^-(x))\times T(Q^+(x))}
\STATE{Find $k_1$ such that\CUT $U(q_1^-,q_1^+)=\alpha_{q_1^-}\otimes k_1\otimes \beta_{q_1^+}$}
\IF{k_1 \mbox{ does not exist}}
\RETURN{False}
\FI
\DOFOREACH{(q^-,q^+)\in Q^-(x)\times Q^+(x)}
\RCOM{{\bf quasi-$\epsilon$}}
\IF {(q^-,q^+) \not\in B(Q^-(x))\times T(Q^+(x))}
\STATE{Find $k$ such that\CUT $U(q^-,q^+)=\alpha_{q^-}\otimes k\otimes \beta_{q^+}$}
\IF{k \mbox{ does not exist}}
\RETURN{False}
\ELIF{k\neq k_1}
\RETURN{False}
\FI
\FI
\OD
\RCOM{{\bf quasi-$\epsilon$}}
\DOFOREACH{(q^-,q^+)\in B(Q^-(x))\times T(Q^+(x))}
\RCOM{{\bf quasi-$\epsilon$}}
\STATE{Find $\gamma (q^-,q^+)$ such that\CUT $U(q^-,q^+)=\alpha_{q^-}\otimes k\otimes \beta_{q^+}\oplus \gamma (q^-,q^+)$}
\RCOM{{\bf quasi-$\epsilon$}}
\IF{\gamma (q^-,q^+) \mbox{ does not exist}}
\RCOM{{\bf quasi-$\epsilon$}}
\RETURN{False}
\FI
\OD
 \RETURN{True}
\end{algo}
}

\begin{definition}[confluence]
Let $G$ be a $\K$-graph and ${\mathbb I}_G$ the acyclic graph having only one vertex. Let $R_1$ be a sequence of $\K$-rules such that 
$$G\underset{R_1}{\longrightarrow} {\mathbb I}_G$$

$\K$-rules are confluent if for all $\K$-graph
 $G_2$ such that there exists $R_2$  a sequence of $\K$-rules with  $G\underset{R_2}{\longrightarrow} G_2$ then there exists $R'_2$ a sequence of $\K$-rules such that  $$G_2\underset{R'_2}{\longrightarrow} {\mathbb I}_G$$
\end{definition}
For the following, $\K$ is a field or a factorial semiring.
\begin{proposition}\label{confluent}
The $\K$-rules are confluent.
\end{proposition}
{\bf Proof } In order to prove this result, we will show that if there exist two applicable $\K$-rules reducing a Glushkov $\K$-graph, then the order of application does not modify the resulting $\K$-graph. 

Let us denote by $r_{x,y}(G)$ the application of a $\K R_1$, $\K R_2$ or $\K R_3$ rule on the vertices $x$ and $y$ with $y=\emptyset$ for a $\K R_3$ rule.

Let $G=(X,U)$ be a Glushkov $\K$-graph and let $r_{x,y}$ and $r_{z,t}$ be two applicable $\K$-rules on $G$ such that  $\{x,y\}\cap \{z,t\}=\emptyset$ and no edge can be deleted by both rules. Necessarily we have  $r_{x,y}(r_{z,t}(G))=r_{z,t}(r_{x,y}(G))$.

Suppose now that $\{x,y\}\cap \{z,t\}\neq\emptyset$ or one edge is deleted by both rules. We have to consider several cases depending on the rule $r_{x,y}$.
\begin{itemize}
\item[$r_{x,y}$ is a $\K R_1$ rule] In this case $r_{z,t}$ can not delete the edge from $x$ to $y$ and $r_{z,t}$ is necessarily a $\K R_1$-rule with $\{x,y\}\cap \{z,t\}\neq\emptyset$. If $y=z$, as the coefficient does not act on the reduction rule, $r_{x,y}( r_{z,t}(G))= r_{z,t}( r_{x,y}(G))$

\item[$r_{x,y}$ is a $\K R_2$ rule] Consider that $r_{z,t}$ is a $\K R_2$ rule with $y=z$. Using the notations of the $\K R_2$ rule, there exist $\alpha _{q^-}$, $\beta _{q^+}$, $l_x,l_y,r_x,r_y$ such that $U(q^-,x)=\alpha _{q^-}l_x$, $U(q^-,y)=\alpha _{q^-}l_y$, $U(x,q^+)=r_x\beta _{q^+}$ and  $U(y,q^+)=r_y\beta _{q^+}$ with $q^-\in Q^-(x)$, $q^+\in Q^+(x)$, and $l_x=\mbox{gcd}_r(x)$, $l_y=\mbox{gcd}_r(y)$ ($r_x=\mbox{gcd}_l(x)$, $r_y=\mbox{gcd}_l(y)$). By hypothesis, a $\K R_2$ rule can also be applied on the vertices $y$ and $t$. There also exists $\alpha ' _{q^-}$, $\beta' _{q^+}$, $l'_x,l'_t,r'_x,r'_t$ such that $\alpha _{q^-}=\alpha ' _{q^-} l'_x$, $\beta _{q^+}= r'_x\beta '_{q^+}$, $U(q^-,t)=\alpha' _{q^-}l'_t$,  $U(t,q^+)=r'_t\beta' _{q^+}$ ($Q^-(x)=Q^-(t)$ and $Q^+(x)=Q^+(t)$). By construction (Algorithm \ref{algoR2}) 
of $\mbox{gcd}_r(x)$, the left gcd of all $\alpha_{q^-}$ is $\b{1}$. Then, whatever  the order of application of $\K R_2$ rules,  the same decomposition of edges values is obtained. Symetrically a same reasoning is applied for the right part.

Consider now that $r_{z,t}=r_{z,\emptyset}$ is a $\K R_3$ rule. Neither edges from $x$ or $y$ nor edges to $x$ or $y$ can be deleted by $r_{z,\emptyset}$. Then $z=x$ or $z=y$. Let $z=y$. If we successively apply $r_{x,y}$ and $r_{y,\emptyset}$  or $r_{y,\emptyset}$ and $r_{x,y}$ on $G$, we obtain the same $\K$-graph following the same method (function {\sc EquivalenceChecking})
 as the previous case. If we choose $z=x$, we have also the same $\K$-graph (commutativity property of the sum operator). 
\item[$r_{x,\emptyset}$ is a $\K R_3$ rule] The only case to consider now is $r_{z,t}=r_{z,\emptyset}$ a $\K R_3$ rule. Suppose that $r_{z,\emptyset}$ deletes an edge also deleted by $r_{x,\emptyset}$ (with $x\neq z$). Let $(q^-,q^+)$ be this edge.

Using the notations of the $\K R_3$ rule, there exist $\alpha _{q^-}$, $\beta _{q^+}$, $l,r$ such that $U(q^-,x)=\alpha _{q^-}l$, $U(x,q^+)=r\beta _{q^+}$, $U(q^-,q^+)=\alpha _{q^-} k \beta _{q^+} \oplus \gamma$ with $q^-\in Q^-(x)$, $q^+\in Q^+(x)$ and $l=\mbox{gcd}_r(x)$, $r=\mbox{gcd}_l(x)$.
 There also exists $\alpha' _{q^-}$, $\beta' _{q^+}$, $l',r'$ such that $U(q^-,z)=\alpha' _{q^-} l'$, $U(z,q^+)=r'\beta' _{q^+}$, $U(q^-,q^+)=\alpha' _{q^-} k' \beta' _{q^+} \oplus \gamma'$ with $l'=\mbox{gcd}_r(z)$, $r'=\mbox{gcd}_l(z)$. By construction, (function {\sc EquivalenceChecking}),
  the computation of  $l$ and $l'$ ($r$ and $r'$) are independant.  A same reasoning is applied for the right part. Then we can choose $\gamma ''$ such that $\gamma=\alpha' _{q^-} k' \beta' _{q^+} \oplus \gamma''$ and $\gamma'=\alpha_{q^-} k \beta _{q^+} \oplus \gamma''$. So 
$ U(q^-,q^+)=\alpha_{q^-} k \beta _{q^+} \oplus \alpha' _{q^-} k' \beta' _{q^+} \oplus \gamma''$. It is easy to see that $r_{x,\emptyset}(r_{z,\emptyset}(G))=r_{z,\emptyset}(r_{x,\emptyset}(G))$.
\end{itemize}
\cqfd

\subsection{$\K$-reducibility}
\begin{definition}
A $\K$-graph $G=(X,U)$ is  said to be {\it $\K$-reducible} if it has no orbit and if it can be  reduced to one vertex by  iterated applications of any of the three rules $\K R_1$, $\K R_2$, $\K R_3$ described below.
\end{definition}


Proposition \ref{K-red}  shows the existence of a sequel of $\K$-rules leading to the complete reduction of Glushkov $\K$-graphs. However, the existence of an algorithm allowing us to obtain this sequel of $\K$-rules depends on the semiring $\K$.

In order to show the $\K$-reducibility property of a Glushkov $\K$-graph $G$, we check (Lemma \ref{2R1}) that every sequence ${\cal R}$ of $\K$-rules leading to the $\K$-reduction of $G$ contains necessarily two $\K R_1$ rules which will be denoted by $r_\circ$ and $r_\bullet$.
\begin{lemma}\label{2R1}
Let $G=(X,U)$ be a $\K$-reducible Glushkov $\K$-graph without orbit with $|X|\geq 3$, and let ${\cal R}=r_1\cdots r_n$ be the sequence of $\K$-rules which can be applied on $G$ and reduce it. Necessarily, ${\cal R}$ can be written ${\cal R'}r_\circ r_\bullet$ with $r_\circ$ and  $r_\bullet$ two $\K R_1$-rules merging respectively $s_I$ and $\Phi$.
\end{lemma}
{\bf Proof }We show this lemma by induction on the number of vertices of the graph. It is obvious that if $|X|=3$ then, the only possible graphs are the following ones:

\VCDraw[.35]{
\begin{VCPicture}{(-15,-2)(15,8)}
\State[s_I]{(0,0)}{0}
\State[\Phi]{(12,0)}{fi}
\State[x]{(6,6)}{x}

\EdgeL{0}{x}{}\LabelL[.5]{\lambda}
\EdgeR{x}{fi}{}\LabelL[.5]{\lambda '}

\State[s_I]{(18,0)}{0}
\State[\Phi]{(30,0)}{fi}
\State[x]{(24,6)}{x}

\EdgeL{0}{x}{}\LabelL[.5]{\lambda}
\EdgeR{x}{fi}{}\LabelL[.5]{\lambda '}
\EdgeR{0}{fi}{}\LabelR[.5]{\lambda ''}

\end{VCPicture}
}

\noindent
and then, for the first one ${\cal R}=r_\circ r_\bullet$ with $k=\lambda$ in $r_\circ$ and $k=\lambda '$ in $r_\bullet$. For the second one $x$ is $\epsilon$-equivalent and ${\cal R}=r r_\circ r_\bullet$ with $r$ a $\K R_3$-rule such that  $\alpha =\b{1}$, $\beta =\b{1}$ , $l=\lambda$, $r=\lambda '$ and $k=\lambda ''$. Then,  $r_\circ$ and $r_\bullet$ are $\K R_1$ rules such that  $k=\b{1}$ for $r_\circ$ and $r_\bullet$. Suppose now that $G$ has $n$ vertices. As it is $\K$-reducible, there exists a sequence of $\K$-rules which leads to one of the two previous basic cases.\cqfd 
 
 For the reduction process, we associate each vertex of $G$ to a
subexpression. We define $E(x)$ to be the expression of the vertex $x$.
At the beginning of the process, $E(x)$ is $a$, the only letter
labelling edges reaching the vertex $x$ (homogeneity of Glushkov
automata). For the vertices $s_I$ and $\Phi$, we define
$E(s_I)=E(\Phi)=\epsilon$. When applying $\K$-rules, we associate a new
expression to each new vertex.  
With notations of figure \ref{KR1-rule}, the $\K R_1$-rule induces
$E(x)\leftarrow E(x)\cdotp k\times E(y)$ with $k=U(x,y)$.
With notations of figure \ref{KR2-rule}, the $\K R_2$-rule induces
$E(x)\leftarrow l_x E(x)r_x + l_yE(y)r_y$.
And with notations of figures \ref{KR3a-rule} and \ref{KR3b-rule}, the
$\K R_3$-rule induces $E(x)\leftarrow l E(x)r + k$.

\begin{proposition}\label{K-red}
Let $G=(X,U)$ be a $\K$-graph without orbit. The graph $G$ is a Glushkov $\K$-graph if and only if it is $\K$-reducible.
\end{proposition}
\noindent
{\bf ( $\Rightarrow$ )} This proposition will be proved by recurrence on the length of the expression. First for $||E||=1$, we have only two proper $\K$-expressions which are $E=\lambda$ and $E=\lambda a \lambda '$, for $\lambda, \lambda '\in \K$. When $E=\lambda$, the Glushkov $\K$-graph has only two vertices which are $s_I$ and $\Phi$ and the edge $(s_I,\Phi )$ is labeled with $\lambda$. Then the $\K R_1$ rule can be applied. Suppose now that $E=\lambda a \lambda '$, then the Glushkov $\K$-graph of $E$ has three vertices and is $\K$-reducible. Indeed, the $\K R_1$-rule can be applied twice.

Suppose now that for each proper $\K$-expression $E$ of length $n$, its Glushkov $\K$-graph is $\K$-reducible. We then have to show that the Glushkov $\K$-graph of $\K$-expressions $F=E+\lambda$, $F=E+\lambda a \lambda '$, $F=\lambda a \lambda '\cdotp E$ and $F=E\cdotp \lambda a \lambda '$ of length $n+1$ are $\K$-reducible. Let us denote by ${\cal R}$ (respectively ${\cal R'}$) the sequence of rules which can be applied on $A _\K (E)$ (respectively $A _\K (F)$). In case $|X|\geq 3$, ${\cal R}={\cal R}_br_\circ r_\bullet$ (respectively ${\cal R'}={\cal R}_b'r'_\circ r'_\bullet$).

\begin{itemize}
\item[case $F=E+\lambda$] We have $\P{F}=\P{E}$, $\F{F}=\F{E}$, $\L{F}=\L{E}$, $\Nl{F}=\Nl{E}+\lambda$ and $\forall i\in \P{E}$, $\Fl{F}=\Fl{E}$. Every rule which can be applied on $A _\K (E)$ and which does not modify the edge $(s_I,\Phi)$ can also be applied on $A _\K (F)$. 

If $A _\K (E)$ has only two states, then ${\cal R}= r$ a $\K R_1$-rule, and then ${\cal R'}= r'$ a $\K R_1$- rule where $r'$ is such that $k=\Nl{E}+\lambda$. Elsewhere, the $(s_I,\Phi)$ edge can only be reduced by a $\K R_3$ rule. 

Suppose now that  there is no $\K R_3$ rule modifying $(s_I,\Phi)$ which can be applied on $A _\K (E)$. Then there is a $\K R_3$ rule $r'$ which can be applied on $A _\K (F)$ with $k=\lambda$ and then  $A _\K (F)$ can be reduced by ${\cal R'}={\cal R}_b r'r'_\circ r'_\bullet$.

Let us now suppose that $r_1,r_2,\cdots r_n$ is the subsequence of $\K R_3$-rules of ${\cal R}$ which modify the $(s_I,\Phi)$ edge. Necessarily, $r_n$ acts on a state $x$ which is $\epsilon$-equivalent. If $Q^-(x)\neq \{s_I\}$ or $Q^+(x)\neq \{\Phi \}$ then ${\cal R}_b'={\cal R}_br_{n+1}$ where $r_n$ in ${\cal R}_b'$ is modified  as follows: $x$ is quasi-$\epsilon$-equivalent with $\gamma=\lambda$ and the rule $r_{n+1}$ is a $\K R_3$ rule on a state $x$ which is $\epsilon$-equivalent and $k=\lambda$. Elsewhere, there is two cases to distinguish. If $\Nl{E}\oplus \lambda=\b{0}$ then the $r_n$ rule is no more applicable on $A _\K (F)$ (no edge between $s_I$ and $\Phi$) and the $r_{n-1}$ rule in ${\cal R}'$ now acts on an  $\epsilon$-equivalent  vertex in $A _\K (F)$. If $\Nl{E}+\lambda\neq \b{0}$ then $r_n$ can be applied on $A _\K (F)$ with $k=k\oplus \lambda$.
\item[case $F=E+\lambda a \lambda '$] If $|\P{E}|=n$, we have, $\P{F}=\P{E}\cup \{n+1\}$, $\F{F}=\F{E}\uplus \{(\lambda,n+1)\}$, $\L{F}=\L{E}\uplus \{(\lambda',n+1)\}$, $\Nl{F}=\Nl{E}$ and $\forall i\in \P{E}$, $\Fl{F}=\Fl{E}$ and $\Fol{F}{n+1}=\emptyset$. In this case, ${\cal R'}={\cal R}_brr'_\circ r'_\bullet$ where $r$ is a $\K R_2$ rule with $\alpha_{s_I}=\beta_{\Phi}=\b{1}$ and $l_y=\lambda$, $r_y=\lambda '$ and so $A _\K (F)$ is $\K$-reducible.
\item[case $F=E\cdotp\lambda a \lambda '$]  If $|\P{E}|=n$, we have, $\P{F}=\P{E}\cup \{n+1\}$, $\F{F}=\F{E}$, $\L{F}=\{(\lambda',n+1)\}$, $\Nl{F}=\emptyset $ and $\forall i\in \P{E}\setminus P(\L{E})$, $\Fl{F}=\Fl{E}$ and $\forall i\in P(\L{E})$, $\Fol{F}{i}=\Fol{E}{i} \uplus \{(\lambda,n+1)\}$.
Let $r_1, \cdots r_n$ be the subsequel of $\K$-rules modifying edges reaching $\Phi$. Necessarily, $n=1$ and $r_1=r_\bullet$ (Lemma \ref{2R1}).
Indeed, let us suppose that $n>1$ and that there exists $j \neq i$ such that $r_j$ is a $\K R_1$, $\K R_2$, or $\K R_3$-rule. Necessarily  $|Q^-(\Phi)|\geq 1$, which contradicts our hypothesis. Then we have ${\cal R}'={\cal R} r_{n+1}$ where $r_\bullet$ the $\K R_1$-rule from a vertex $x$ to $\Phi$  of the sequence ${\cal R}$ and labeled with $k_i$ is modified in ${\cal R'}$ as follows: $k = k_i\otimes \lambda$. We have also $k =\lambda '$ for the rule $r_{n+1}$.

\end{itemize}
The case $F=  \lambda a \lambda '\cdotp E$ is proved similarily as the previous one considering the rules modifying edges from $s_I$ (with $r_\circ$ instead of $r_\bullet$). \\

\noindent
{\bf ( $\Leftarrow$ )} By induction on the number of states of the reducible $\K$-graph $G=(X,U)$.
If $|X|=2$, $X=\{s_I,\Phi\}$ and the only $\K$-expression $E$ is
$\lambda$ with $\lambda \in \K$. Let $G'=(X',U')$ be the Glushkov
$\K$-graph obtained from $E$. By construction
$\lambda=U(s_I,\Phi)=E(s_I)$ and $\lambda=\Nl{E}$, necessarily $G'=G$.

We consider the property true for ranks bellow $n+1$ and $G$ a
$\K$-graph partially reduced.
Three cases can occur according to the graphic form of the partially
reduced graph.
Either we will have to apply twice the $\K R_1$-rule or once the $\K
R_3$-rule and twice the $\K R_1$-rule if $X=\{s_I,x,\Phi\}$, or we will
have to apply once the $\K R_2$-rule and twice the $\K R_1$-rule if
$X=\{s_I,x,y,\Phi\}$. For each case, we compute successively the new
expressions of vertices, and we check that the Glushkov
construction applied on the final $\K$-expression is $G$.
\cqfd

\subsection{Several examples of use for $\K$-rules}
For the $\K R_2$ rule, the first example is for transducers in $(\K, \oplus, \otimes)=$($\Sigma ^* \cup \emptyset , \cup, \cdotp$) where ``$\cdotp$'' denotes the concatenation operator. In this case, we can express the $\K R_2$ rule conditions as follows. For all $q^-$ in $Q^-(x)$,  $\alpha_{q^-}$ is the  common prefix of $U(q^-,x)$ and $U(q^-,y)$. Likewise, for all $q^+$ in $Q^+(x)$, $\beta_{q^+}$ is the  common suffix of $U(q^+,x)$ and $U(q^+,y)$ .

\medskip
\VCDraw[.85]{
\begin{VCPicture}{(-3,-2)(20,2)}
\State[p_2]{(0,-1.5)}{p2}
\State[p_1]{(0,1.5)}{p1}
\State[y]{(4,-1.5)}{y}
\State[x]{(4,1.5)}{x}
\State[q_2]{(8,-1.5)}{q2}
\State[q_1]{(8,1.5)}{q1}
\EdgeL{p1}{x}{}\LabelL[.5]{aa}
\EdgeR{p1}{y}{}\LabelL[.3]{ab}
\EdgeR{p2}{y}{}\LabelR[.6]{b}
\EdgeL{p2}{x}{}\LabelL[.2]{a}
\EdgeL{x}{q1}{}\LabelL[.5]{aba}
\EdgeL{x}{q2}{}\LabelL[.3]{aa}
\Edge{y}{q1}{}\LabelL[.2]{bba}
\Edge{y}{q2}{}\LabelR[.5]{ba}
\State[p_2]{(12,-1.5)}{p2a}
\State[p_1]{(12,1.5)}{p1a}
\StateVar[a E(x) a + b E(y) b]{(17,0)}{xy}
\State[q_2]{(22,-1.5)}{q2a}
\State[q_1]{(22,1.5)}{q1a}
\EdgeL{p1a}{xy}{}\LabelL[.3]{a}
\EdgeL{p2a}{xy}{}\LabelL[.3]{\varepsilon}
\EdgeL{xy}{q1a}{}\LabelL[.7]{ba}
\EdgeL{xy}{q2a}{}\LabelL[.7]{a}

\end{VCPicture}
}
\medskip

The second one is in ($\Z/ 7\Z [i,j,k],\oplus,\otimes$), where $\{i,j,k\}$ are elements of the quaternions and $\oplus$ is the sum and $\otimes$ the product.  In this case, $\K$ is a field. Every factorization leads to the result.

\medskip
\VCDraw[.85]{
\begin{VCPicture}{(-3,-2)(20,2)}
\State[p_2]{(0,-1.5)}{p2}
\State[p_1]{(0,1.5)}{p1}
\State[y]{(4,-1.5)}{y}
\State[x]{(4,1.5)}{x}
\State[q_2]{(8,-1.5)}{q2}
\State[q_1]{(8,1.5)}{q1}
\EdgeL{p1}{x}{}\LabelL[.5]{2i}
\EdgeR{p1}{y}{}\LabelL[.3]{j}
\EdgeR{p2}{y}{}\LabelR[.6]{-k}
\EdgeL{p2}{x}{}\LabelL[.2]{2}
\EdgeL{x}{q1}{}\LabelL[.5]{3j}
\EdgeL{x}{q2}{}\LabelL[.3]{j}
\Edge{y}{q1}{}\LabelL[.2]{2k}
\Edge{y}{q2}{}\LabelR[.5]{2k}
\State[p_2]{(12,-1.5)}{p2a}
\State[p_1]{(12,1.5)}{p1a}
\StateVar[2i E(x) 1 + j E(y) 2i]{(17,0)}{xy}
\State[q_2]{(22,-1.5)}{q2a}
\State[q_1]{(22,1.5)}{q1a}
\EdgeL{p1a}{xy}{}\LabelL[.3]{1}
\EdgeL{p2a}{xy}{}\LabelL[.3]{-i}
\EdgeL{xy}{q1a}{}\LabelL[.7]{3j}
\EdgeL{xy}{q2a}{}\LabelL[.7]{j}

\end{VCPicture}
}
\medskip

We now give a complete example using the three rules on the ($\N \cup \{+\infty \},min,+$) semiring. This example enlightens the reader on the problem of the quasi-epsilon equivalence. For this example, we will identify the vertex with its label.

\medskip
\VCDraw[.50]{
\begin{VCPicture}{(0,-6)(40,5)}
\State[s_I]{(0,0)}{0}
\State[y]{(8,0)}{y}
\State[x]{(8,4)}{x}
\State[z]{(8,-3)}{z}
\State[\Phi]{(18,0)}{fi}
\EdgeL{0}{x}{}\LabelL[.5]{2}
\EdgeR{0}{y}{}\LabelL[.6]{6}
\EdgeR{0}{z}{}\LabelL[.6]{2}
\VArcR{arcangle=-50,ncurv=.8}{0}{fi}{3}
\EdgeR{x}{y}{}\LabelR[.6]{5}
\EdgeL{x}{fi}{}\LabelL[.6]{6}

\Edge{y}{fi}{}\LabelL[.35]{2}
\Edge{z}{fi}{}\LabelR[.5]{0}

\State[s_I]{(24,0)}{0a}
\State[y]{(32,0)}{ya}
\State[x]{(32,4)}{xa}
\State[z]{(32,-3)}{za}
\State[\Phi]{(42,0)}{fia}
\EdgeL{0a}{xa}{}\LabelL[.5]{0\otimes 2}
\EdgeR{0a}{ya}{}\LabelL[.6]{0\otimes 6\otimes 0}
\EdgeR{0a}{za}{}\LabelL[.6]{2}
\VArcR{arcangle=-50,ncurv=.8}{0a}{fia}{0\otimes 6\otimes 1\oplus 3}
\EdgeR{xa}{ya}{}\LabelL[.6]{5\otimes 0}
\EdgeL{xa}{fia}{}\LabelL[.6]{5\otimes 1}

\Edge{ya}{fia}{}\LabelL[.35]{2}
\Edge{za}{fia}{}\LabelR[.5]{0}

\rput[lt](23,-7.2){\Large $\K R_3$ rule can be applied on $x$ with $l=2$, $r=5$ and $k=6$}
\end{VCPicture}
}

\VCDraw[.50]{
\begin{VCPicture}{(0,-8)(40,9)}
\State[s_I]{(0,0)}{0}
\State[y]{(8,0)}{y}
\StateVar[2 x 5 + 6]{(8,4)}{x}
\State[z]{(8,-3)}{z}
\State[\Phi]{(18,0)}{fi}
\EdgeL{0}{x}{}\LabelL[.5]{0}
\EdgeL{x}{fi}{}\LabelL[.5]{0\otimes 1\otimes 0}
\EdgeR{0}{z}{}\LabelR[.6]{2}
\VArcR{arcangle=-50,ncurv=.8}{0}{fi}{3}
\EdgeR{x}{y}{}\LabelR[.6]{0\otimes 0}

\Edge{y}{fi}{}\LabelL[.35]{2\otimes 0}
\Edge{z}{fi}{}\LabelR[.5]{0}

\State[s_I]{(24,0)}{0a}
\StateVar[0 y 2 + 1]{(32,0)}{ya}
\StateVar[2 x 5 + 6]{(32,4)}{xa}
\State[z]{(32,-3)}{za}
\State[\Phi]{(42,0)}{fia}
\EdgeL{0a}{xa}{}\LabelL[.5]{0}
\EdgeR{0a}{za}{}\LabelR[.6]{2}
\VArcR{arcangle=-50,ncurv=.8}{0a}{fia}{3}
\EdgeR{xa}{ya}{}\LabelR[.6]{0}

\Edge{ya}{fia}{}\LabelL[.35]{0}
\Edge{za}{fia}{}\LabelR[.5]{0}

\rput[lt](0,-7.2){\Large $\K R_3$ rule can be applied on $y$}
\rput[lt](0,-8.2){\Large with $l=0$, $r=2$ and $k=1$}
\rput[lt](24,-7.2){\Large $\K R_1$ rule can be applied on $(2x5+6)$ and on $(0y2+1)$}

\end{VCPicture}
}

\VCDraw[.75]{
\begin{VCPicture}{(0,-6)(20,5)}
\State[s_I]{(0,-1)}{0a}
\State[z]{(6,-1)}{za}
\State[\Phi]{(12,-1)}{fia}
\LargeState \StateVar[(2 x 5 + 6)(0 y 2 + 1)]{(6,2)}{xa}
\EdgeL{0a}{xa}{}\LabelL[.3]{0}
\EdgeR{0a}{za}{}\LabelR[.5]{2}
\VArcR{arcangle=-30,ncurv=.8}{0a}{fia}{3}
\EdgeL{xa}{fia}{}\LabelL[.8]{0}
\Edge{za}{fia}{}\LabelR[.5]{0}
\MediumState
\State[s_I]{(15,-1)}{0}
\State[\Phi]{(31,-1)}{fi}
\LargeState \StateVar[(2 x 5 + 6)(0 y 2 + 1) + 2 z]{(23,2)}{x}
\EdgeL{0}{x}{}\LabelL[.5]{0}
\EdgeL{0}{fi}{3}

\EdgeL{x}{fi}{}\LabelL[.5]{0}

\rput[lt](0,-4.2){\Large $\K R_2$ rule can be applied on}
\rput[lt](0,-5.2){\Large $(2x5+6)(0y2+1)$ and on $z$}
\rput[lt](15,-3.2){\Large A $\K R_3$ rule can be applied to end the process}

\end{VCPicture}
}

\bigskip

\noindent
This example leads to a possible $\K$-expression such as $E=((2x5+6)(0y2+1)+2z)+3$

\section{Glushkov $\K$-graph with orbits}

We will now consider a graph which has at least one maximal orbit $\O$. We extend the notions of strong stability and strong transversality to the $\K$-graphs obtained from $\K$-expressions in SNF.
We have to give a characterization on coefficients only. 
The stability and transversality notions are rather linked. Indeed, if we consider the states of $\E{\O}$ as those of $\O ^+$ then both notions amount to the transversality. Moreover, the extension of these notions to WFAs ($\K$-stability - definition \ref{Kstable} - and $\K$-transversality - definition \ref{Ktransverse}), implies the manipulation of output and input vectors of $\O$ whose product is exactly the orbit matrix of $\O$ (Proposition \ref{Kbal}). 

\begin{lemma}\label{orbit=star}
Let $E$ be a $\K$-expression and $G_\K(E)$ its Glushkov $\K$-graph. Let $\O=(X_\O,U_\O)$ be a maximal orbit of $G_\K(E)$. Then $E$ contains a closure subexpression $F$ such that $X_\O=\P{F}$.
\end{lemma}  
This lemma is a direct consequence of Lemma 4.5 in \cite{CZ97} and of Lemma \ref{rat1}.
\begin{definition}[$\K$-stability]\label{Kstable}
A maximal orbit $\O$ of a $\K$-graph $G=(X,U)$ is $\K$-stable if 
\begin{itemize} 
\item ${\widetilde \O}$ is stable and
\item the matrix $M_{\O} \in \K^{|\S{\O}| \times|\E{\O}|}$ such that
$M_{\O}(s,e)=U(s,e)$, for each $(s,e)$ of $\S{\O} \times \E{\O}$,  can be written
as a product $VW$ of two vectors such that $V \in \K^{ |\S{\O}| \times
1}$ and $W \in \K^{1 \times |\E{\O}|}$.
\end{itemize}
The graph $G$ is $\K$-stable if each of its maximal orbits is $\K$-stable.
\end{definition}

If a maximal orbit $\O$ is $\K$-stable, $M_{\O}$ is a matrix of rank $1$ called the {\it orbit matrix}. Then, for a decomposition of $M_{\O}$ in the product $VW$ of two vectors, $V$ will be called the {\it tail-orbit vector} of $\O$ and $W$ will be called the {\it head-orbit vector} of $\O$. 
\begin{lemma}\label{lmKstable}
A Glushkov $\K$-graph obtained from a $\K$-expression $E$ in SNF is $\K$-stable.
\end{lemma}
{\bf Proof }
Let $G$ be the Glushkov $\K$-graph of a $\K$-expression $E$ in SNF, $A_\K(E)=(\Sigma,Q,s_I,F,\delta)$ its Glushkov WFA  and $\O=(X_\O,U_\O)$ be a maximal orbit of G. 
Following Lemma \ref{rat1} and Theorem \ref{th}, $G$ is strongly stable which implies that every orbit of $G$ is stable. 
Let $s_i\in \S{\O}$, $1\leq i\leq |\S{\O}|$ and $e_j\in \E{\O}$, $1\leq j\leq \S{\O}$. Following the extended Glushkov construction and as for all $s_i\in \S{\O}$, $s_i\neq s_I$, we have $\delta (s_i,a,e_j)= \mbox{Coeff}_{\Fol{E}{s_i}}(e_j)$. As $\O$ corresponds to a closure subexpression $F^*$ or $F^+$ (Lemma \ref{orbit=star}) and as $(s_i,a,e_j)$ is an edge of $X_\O\times \Sigma \times X_\O$, we have $\delta (s_i,a,e_j)= \mbox{Coeff}_{\Fol{F^*}{s_i}}(e_j)= \mbox{Coeff}_{\Fol{F}{s_i}\uplus  \mbox{{\small Coeff}}_{\L{F}}(s_i).\F{F}}(e_j)$. As $E$ is in SNF, so are $F^*$ and $F^+$,  and then $\delta (s_i,a,e_j)= \mbox{Coeff}_{\mbox{{\small Coeff}}_{\L{F}}(s_i).\F{F}}(e_j)= \mbox{Coeff}_{\L{F}}(s_i).\mbox{Coeff}_{\F{F}}(e_j)$. The lemma is proved  choosing $V\in \K^{ |\S{\O}| \times 1}$ such that $V(i,1)=\mbox{Coeff}_{\L{F}}(s_i)$  and $W\in \K^{1 \times |\E{\O}|}$ with $W(1,j)=\mbox{Coeff}_{\F{F}}(e_j)$.
\cqfd

\begin{definition}[$\K$-transversality]\label{Ktransverse}
A maximal orbit $\O$ of $G=(X,U)$ is $\K$-transverse if 
\begin{itemize} 
\item  $\widetilde{\O}$ is transverse,
\item the matrix $M_e \in \K^{|\O^-| \times
|\E{\O}|}$ such that $M_e(p,e)=U(p,e)$  for each $(p,e)$ of $\O^-\times \E{\O}$, can be written
as a product $ZT$ of two vectors such that $Z \in \K^{|\O^-| \times 1}$ and
$T\in \K^{1 \times |\E{\O}|}$,
\item the matrix $M_s \in \K^{|\S{\O}| \times |\O^+|}$ such that
$M_s(s,q)=U(s,q)$ for each $(s,q)$ of $\S{\O}\times \O^+$, can be written
as a product $T'Z'$ of two vectors such that $T' \in \K^{ |\S{\O}| \times
1}$ and $Z' \in \K^{1 \times |\O^+|}$.
\end{itemize}
The graph $G$ is $\K$-transverse if each of its maximal orbits is $\K$-transverse.
\end{definition}

If a maximal orbit $\O$ is $\K$-transverse, $M_e$ (respectively $M_s$) is a matrix of rank $1$ called the {\it input matrix} of $\O$ (respectively {\it output matrix} of $\O$). For a decomposition of $M_e$ (respectively $M_s$) in the product $ZT$ (respectively $T' Z'$) of two vectors, $T$ will be called the {\it input vector} (respectively $T'$ will be called the {\it output vector}) of $\O$. 

\begin{lemma}\label{lmKtransverse}
The Glushkov $\K$-graph $G=(X,U)$ of a $\K$-expression $E$ in SNF is $\K$-transverse.
\end{lemma}
{\bf Proof }
Let $\O$ be a maximal orbit of G.  Following Lemma \ref{rat1} and Theorem \ref{th}, $G$ is strongly transverse implies that $\O$ is transverse. By Lemma  \ref{orbit=star}, there exists a maximal closure subexpression $H$ such that $H=F^*$ or $H=F^+$. As $E$ is in SNF, so is $H$. By the definition of the function {\it Follow}, we have in this case: for all $p\in \S{\O}$, for all $q\in O^+$, $U(p,q)=\mbox{Coeff}_{\Fol{F}{p} }(q)$. We now have to distinguish three cases. 
\begin{enumerate}
\item If $|\O^+|=1$, then the result holds immediatly. Indeed the output matrix of $\O$ is a vector.
\item If $\O^+=\{q_1,\cdots ,q_n\}$ and $n>1$, $\forall 1\leq j\leq n, q_j\neq \Phi$, necessarily, we have $\O^+=\displaystyle{\bigcup_l P(\F{H_l})}$ with $H_l$ some subexpressions of $E$. Then we have $U(p,q_j)= \mbox{Coeff}_{\mbox{Coeff}_{\L{F}}(p).\F{H_l}}(q_j)$ if $q_j\in P(\F{H_l})$. Then as $q_j$ is a first position of only one subexpression,  $U(p,q_j)=k_p\otimes \mbox{Coeff}_{\F{H_l}}(q_j)$ where $k_p=\mbox{Coeff}_{\L{F}}(p)$ which concludes this case.
\item Now if $\exists 1\leq j\leq n \mid q_j=\Phi$ then $U(p,q_j)=\mbox{Coeff}_{\L{F}}(p)\otimes k$ where $k$ is the $Null$ value of some subexpression following $F$ not depending on $p$.
\end{enumerate}
 A same reasoning can be used for the left part of the transversality. \cqfd 
 
 \begin{definition}[$\K$-balanced]
The orbit $\O$ of a graph $G$ is $\K$-balanced if  $G$ is $\K$-stable and  $\K$-transverse and if there exists  an input vector $T$ of $\O$ and  an output vector $T'$ of $\O$ such that the orbit matrix $M_\O=T'T$. The graph $G$ is $\K$-balanced if every maximal orbit of $G$ is $\K$-balanced.
\end{definition}

\begin{proposition}\label{Kbal}
A Glushkov $\K$-graph obtained from a $\K$-expression $E$ in SNF is $\K$-balanced. 
\end{proposition}
{\bf Proof }
 Lemma  \ref{lmKstable} enlightens on the fact that $V$, the tail orbit vector of $\O$, is  such that $V(i,1)=\mbox{Coeff}_{\L{F}}(i)$ for all $i\in P(\L{F})$, which is, from Lemma  \ref{lmKtransverse}, the output vector of $\O$.
The details of the proofs for these lemmas show in the same way that there exists an  head-orbit vector and an  input vector for  $\O$  which are equal.
\cqfd

We can now define the recursive version of WFA $\K$-balanced property.

\begin{definition}
A $\K$-graph is strongly $\K$-balanced  if (1) it has no orbit or (2) it is $\K$-balanced and if after deleting all edges $\S{\O}\times \E{\O}$ of each maximal orbit $\O$, it is strongly $\K$-balanced.
\end{definition}

\begin{proposition}\label{K-prop}
A Glushkov $\K$-graph obtained from a $\K$-expression $E$ in SNF is strongly $\K$-balanced.
\end{proposition}
{\bf Proof }
Let $G$ be the Glushkov of a $\K$-expression $E$ and $\O$ be a maximal orbit of G. The Glushkov $\K$-graph $G$ is strongly stable and strongly transverse. As $E$ is in $SNF$, edges of $\S{\O}\times \E{\O}$ that are deleted are backward edges of a unique closure subexpression $F^*$ or $F^+$. Consequently, the recursive process of edges removal deduced from the definition of strong $\K$-stability produces only maximal orbits which are $\K$-balanced. The orbit $\O$ is therefore strongly $\K$-balanced.
\cqfd

\begin{theorem}
Let $G=(X,U)$. $G$ is a Glushkov $\K$-graph of a $\K$-expression $E$ in SNF if and only if
\begin{itemize}
\item $G$ is strongly $\K$-balanced.
\item The graph without orbit of $G$ is $\K$-reducible.
\end{itemize}
\end{theorem}
{\bf Proof } Let $G=(X,U)$ be a Glushkov $\K$-graph. From Proposition \ref{K-prop}, $G$ is strongly $\K$-balanced. The graph without orbit of $G$ is $\K$-reducible (Proposition \ref{K-red})
For the converse part of the theorem, if $G$ has no orbit and $G$ is $\K$-reducible, by Proposition  \ref{K-red} the result holds immediatly. Let $\O$ be a maximal orbit of $G$. As it is strongly $\K$-balanced, we can write $M_\O=VW$ the orbit matrix of $\O$, there exists an output vector $T'$ equal to the tail-orbit vector $V$ and an input vector $T$ equal to the head-orbit vector $W$. If the graph without orbit of $\O$ corresponds to a $\K$-expression $F$ then $\O$ corresponds to the $\K$-expression $F^+$ where $\mbox{Coeff}_{\F{F^+}}(i)=W(1,i), \forall i \in P(\F{F^+})$,  $\mbox{Coeff}_{\L{F^+}}(j)=V(j,1), \forall j \in P(\L{F^+})$. We have also $\mbox{Coeff}_{\Fol{F^+}{j}}(i)=\mbox{Coeff}_{\Fol{F}{j}\uplus \mbox{{\small Coeff}}_{\L{F}}.\F{F}}(i)$, $\forall j\in P(\L{F})$ and $\forall i\in P(\F{F})$. Hence the Glushkov functions are well defined. 

We now have to show that the graph without orbit of $\O$ can be reduced to a single vertex. By the successive applications of the $\K$-rules, the vertices of the graph without orbit of $\O$ can be reduced to a single state (giving a $\K$-rational expression for $\O$). Indeed, as $\O$ is transverse, no $\K$-rule concerning one vertex of $\O$ and one vertex out of $\O$ can be applied. 

\cqfd

\section{Algorithm for orbit reduction}
In this section, we present a recursive algorithm that computes a $\K$-expression from a Glushkov $\K$-graph. We then give an example which illustrate this method.
\subsection*{Algorithms}
{\small 
\begin{algorithm}[H]
\begin{algo}{OrbitReduction}{G}
\IN { A $\K$-graph $G=(X,U)$}
\OUT {A newly computed graph without orbit}

\DOFOREACH{\mbox{maximal orbit }\O=(X_\O,U_\O)\mbox{ of }G}
\IF{\CALL{BackEdgesRemoval}{\O,T,T',Z,Z'}}
\IF{\CALL{OrbitReduction}{\O}}
\IF{\CALL{Expression}{E_{\O},\O,T,T'}}
\CALL{ReplaceStates}{G,\O,E_{\O},Z,Z'}
\ELSE
\RETURN{False}
\FI
\ELSE
\RETURN{False}
\FI
\ELSE
\RETURN{False}
\FI
\OD
\RETURN{True}
\end{algo} 
\end{algorithm}
}

The {\sc BackEdgesRemoval} function on $\O$ deletes edges from $\S{\O}$ to $\E{\O}$, returns true if vectors $T,T',Z,Z'$ (as defined in definition \ref{Ktransverse}) can be computed, false otherwise.

The {\sc Expression} function returns true, computes the $\K$-expression $E$ of $G'=(X_{\O} \cup \{s_I,\Phi\},U')$ where $U'\leftarrow U_{\O} \cup \{(s_I,T(1,j),e_j)\mid e_j\in \E{\O}\}\cup \{(s_i,T'(i,1),\Phi)\mid s_i\in \S{\O}\}$ and  ouputs $E_{\O}\leftarrow E^+$ if $\O$ is $\K$-reducible. It returns false otherwise.

The {\sc ReplaceStates} function replaces $\O$ by one state $x$ labeled $E_{\O}$ and connected to $\O^-$ and $\O^+$ with the sets of coefficients of $Z$ and $Z'$. Formally $G=(X\setminus X_\O\cup \{x\},U)$ with $U\leftarrow U\setminus \{(u,k,v)\mid u,v\in \O\}\cup \{(p_j,Z(j,1),x)\mid p_j\in \O^-\}\cup \{(x,Z'(1,i),q_i)\mid q_i\in \O^+\}$.\\

{\small
\begin{algorithm}[H]
\begin{algo}{BackEdgesRemoval}{\O,M_e, M_s,T,T',Z,Z'}
\IN { a $\K$-graph $\O=(X_\O,U_\O)$, $M_e\in \K^{|\O^-|\times|\E{\O}|}$}
\IN{ $M_s\in \K^{|\S{\O}|\times|\O^+|}$}
\OUT{$T\in \K^{1\times|\E{\O}|}$, $T'\in \K^{|\S{\O}|\times 1}$,$Z\in \K^{|\O^-|\times 1}$,$Z'\in \K^{1 \times|\O^+|}$} 

\DOFOREACH {\mbox{line }l\mbox{ of }M_e}
\SET{\mbox{gcd}_l(l)}{\mbox{{\sc left gcd} of all values of the line } l}
\OD
\COM{$\mbox{gcd}_l$ is the vector of $\mbox{gcd}_l(l)$ values} 
\STATE{Find a vector $\b{\mbox{gcd}_l}$ such that $M_e= \mbox{gcd}_l\otimes \b{\mbox{gcd}_l}$}
\IF {\b{\mbox{gcd}_l}\mbox{ does not exist}}
\RETURN{False}
\FI

\DOFOREACH {\mbox{column }c\mbox{ of }M_s}
\SET{\mbox{gcd}_r(c)}{\mbox{{\sc right gcd} of all values of the column } c}
\OD
\COM{$\mbox{gcd}_r$ is the vector of $\mbox{gcd}_r(c)$ values} 
\STATE{Find a vector $\b{\mbox{gcd}_r}$ such that $M_s= \b{\mbox{gcd}_r}\otimes \mbox{gcd}_r$}
\IF {\b{\mbox{gcd}_r}\mbox{ does not exist}}
\RETURN{False}
\FI

\STATE{Find $k$ such that $M_{\O}= \b{\mbox{gcd}_r} \otimes k \otimes \b{\mbox{gcd}_l}$}
\COM{$M_{\O}\in \K^{|\S{\O}|\times|\E{\O}|}$ is the orbit matrix of $\O$}
\IF{k \mbox{ does not exist}}
\RETURN{False}
\FI

\SET{A}{\mbox{{\sc right gcd} of all values of the }\mbox{gcd}_l\mbox{ vector}}
\SET{B}{\mbox{{\sc left gcd} of all values of the }\mbox{gcd}_r\mbox{ vector}}

\SET{k_1}{\CALL{left gcd}{B,k}}
\STATE{Find $k_2$ such that $k=k_1\otimes k_2$} 
\IF{\CALL{right gcd}{k_2,A}\neq k_2}
\RETURN{False}
\FI
\SET{T}{k_2\otimes \b{\mbox{gcd}_l}}
\SET{T'}{\b{\mbox{gcd}_r}\otimes k_1}
\STATE{Find $Z$ such that ${\mbox{gcd}_l}=Z\otimes k_2$}
\STATE{Find $Z'$ such that ${\mbox{gcd}_r}=k_1\otimes Z'$}

\STATE {delete any edge from $\S{\O}$ to $\E{\O}$}

\RETURN{True}
\end{algo}
\end{algorithm}
}

\subsection*{Illustrated example}

We illustrate Glushkov WFAs characteristics developped in this paper  with a reduction example in the $(\N \cup \{+\infty \},min,+)$ semiring. This example deals with the reduction of an orbit and its connection to the outside. We first reduce the orbit to one state and replace the orbit by this state in the original graph. This new state is then linked to the predecessors (respectively successors) of the orbit with vector $Z$ (respectively $Z'$) as label of edges.

\begin{figure}[H]
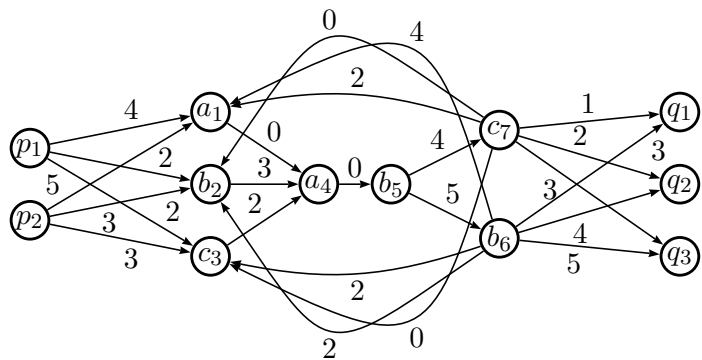

\VCDraw[0.8]{
\begin{VCPicture}{(-8,-4.5)(16,5)}
\State[p_1]{(-2,1)}{p1}
\State[p_2]{(-2,-1)}{p2}

\State[q_1]{(16,2)}{q1}
\State[q_2]{(16,0)}{q2}
\State[q_3]{(16,-2)}{q3}

\State[a_1]{(3,2)}{e1}
\State[b_2]{(3,0)}{e2}
\State[c_3]{(3,-2)}{e3}

\State[b_5]{(8,0)}{y}
\State[a_4]{(6,0)}{x}

\State[b_6]{(11,-1.5)}{s2}
\State[c_7]{(11,1.5)}{s1}


\EdgeL{p1}{e1}{}\LabelL[.6]{4}
\EdgeL{p1}{e2}{}\LabelL[.8]{2}
\EdgeL{p1}{e3}{}\LabelL[.8]{2}
\EdgeL{p2}{e1}{}\LabelL[.1]{5}
\EdgeL{p2}{e2}{}\LabelR[.4]{3}
\EdgeL{p2}{e3}{}\LabelR[.6]{3}

\EdgeL{e1}{x}{}\LabelL[.5]{0}
\EdgeL{e2}{x}{}\LabelL[.5]{3}
\EdgeL{e3}{x}{}\LabelL[.5]{2}

\EdgeL{x}{y}{}\LabelL[.5]{0}

\EdgeL{y}{s1}{}\LabelL[.5]{4}
\EdgeL{y}{s2}{}\LabelL[.5]{5}

\EdgeL{s1}{q1}{}\LabelL[.5]{1}
\EdgeL{s2}{q1}{}\LabelR[.9]{3}
\EdgeL{s1}{q2}{}\LabelL[.4]{2}
\EdgeL{s2}{q2}{}\LabelR[.4]{4}
\EdgeL{s1}{q3}{}\LabelR[.3]{3}
\EdgeL{s2}{q3}{}\LabelR[.4]{5}


\ArcR{s1}{e1}{}\LabelR[.5]{2}
\VArcR{arcangle=-45,ncurv=1.5}{s1}{e2}{}\LabelR[.5]{0}
\VArcR{arcangle=50,ncurv=1.5}{s1}{e3}{}\LabelL[.5]{0}

\VArcL{arcangle=-50,ncurv=1.5}{s2}{e1}{}\LabelR[.5]{4}
\VArcL{arcangle=45,ncurv=1.5}{s2}{e2}{}\LabelL[.5]{2}
\ArcL{s2}{e3}{}\LabelL[.5]{2}
\end{VCPicture}
}
\caption{An example for orbit reduction}\label{fig-example}
\end{figure}

Let $G$ be the $\K$-subgraph of Figure \ref{fig-example} and let $\O$ be the only maximal orbit of $G$ such that $X_\O=\{a_1,b_2,c_3,a_4,b_5,b_6,c_7\}$.

%
%

\medskip

%
%
 
We have $M_s=\left(\begin{array}{ccc}1&2&3\\3&4&5\end{array}\right)$,
$M_e=\left(\begin{array}{ccc}4&2&2\\5&3&3\end{array}\right)$. We can check that $\O$ is $\K$-transverse. $M_s=\left(\begin{array}{c}0\\2\end{array}\right) \left(\begin{array}{ccc}1&2&3\end{array}\right)=T' Z'$ and $M_e=\left(\begin{array}{c}2\\3\end{array}\right) \left(\begin{array}{ccc}2&0&0\end{array}\right)=Z T$.\\ 

We then verify that the orbit $\O$ is $\K$-stable.
$M_\O=\left(\begin{array}{ccc}2&0&0\\4&2&2\end{array}\right)=\left(\begin{array}{c}0\\2\end{array}\right) \left(\begin{array}{ccc}2&0&0\end{array}\right)=VW$.
 We easily check that the orbit is $\K$-balanced. There is an input vector $T$ which is equal to $W$ and an output vector $T'$ which is equal to $V$.\\
 
Then, we delete back edges and add $s_I$ and $\Phi$ vertices for the orbit $\O$. The $s_I$ vertex is connected to $\E{\O}$. Labels of edges are values of the $T$  vector.
Every vertex of $\S{\O}$ is connected to $\Phi$. Labels of edges are values of the $T'$ vector. The following graph is then reduced to one state by iterated applications of $\K$-rules. 

\VCDraw[1]{
\begin{VCPicture}{(-4,-2.5)(16,2.5)}
\State[s_I]{(-1,0)}{pi}
\State[\Phi]{(15,0)}{qj}
\State[a]{(3,2)}{e1}
\State[b]{(3,0)}{e2}
\State[c]{(3,-2)}{e3}

\State[b]{(8,0)}{y}
\State[a]{(6,0)}{x}

\State[c]{(11,-1.5)}{s2}
\State[b]{(11,1.5)}{s1}

\EdgeL{pi}{e1}{}\LabelL[.6]{2}
\EdgeL{pi}{e3}{}\LabelR[.6]{0}
\EdgeL{s2}{qj}{}\LabelR[.4]{2}
\EdgeL{pi}{e2}{}\LabelL[.6]{0}
\EdgeL{s1}{qj}{}\LabelL[.4]{0}


\EdgeL{e1}{x}{}\LabelL[.5]{0}
\EdgeL{e2}{x}{}\LabelL[.5]{3}
\EdgeL{e3}{x}{}\LabelL[.5]{2}

\EdgeL{x}{y}{}\LabelL[.5]{0}

\EdgeL{y}{s1}{}\LabelL[.5]{4}
\EdgeL{y}{s2}{}\LabelL[.5]{5}

\end{VCPicture}
}

 The expression $F$ associated to this graph is replaced by $F^+$ and states of $\O^-$ (respectively $\O^+$) are connected to the newly computed state choosing $Z$ as vector of coefficients (respectively $Z'$).

\VCDraw[1]{
\begin{VCPicture}{(-4,-2)(16,2)}
\State[p_1]{(0,1)}{p1}
\State[p_2]{(0,-1)}{p2}

\State[q_1]{(15,1)}{q1}
\State[q_2]{(15,0)}{q2}
\State[q_3]{(15,-1)}{q3}

\FixStateDiameter{2cm}
\StateVar[\big( (2a+b3+c2)\cdotp a\cdotp  b\cdotp (4b+5c2)\big) ^{\small +}]{(7.5,0)}{e1}
\MediumState

\EdgeL{p1}{e1}{}\LabelL[.5]{2}
\EdgeL{p2}{e1}{}\LabelL[.5]{3}
\EdgeL{e1}{q1}{}\LabelL[.5]{1}
\EdgeL{e1}{q2}{}\LabelL[.5]{2}
\EdgeL{e1}{q3}{}\LabelL[.5]{3}

\end{VCPicture}
}
\medskip

\section{Conclusion}
While trying to characterize Glushkov $\K$-graph, we have pointed out an error in the paper by Caron and Ziadi \cite{CZ97} that we have corrected. This patching allowed us to extend characterization to $\K$-graph restricting $\K$ to  factorial semirings or fields. For fields, conditions of applications of $\K$-rules are sufficient to have an algorithm.\\

For the case of strict semirings,
this limitation allowed us to work with {\sc gcd} and then to give algorithms of computation of $\K$-expressions from Glushkov $\K$-graphs. 

This characterization is divided into two main parts. The first one is the reduction of an acyclic Glushkov $\K$-graph into one single vertex labeled with the whole $\K$-expression. We can be sure that this algorithm ends without doing a depth first search according to confluence of $\K$-rules. The second one is lying on orbit properties. These criterions allow us to give an algorithm computing a single vertex from each orbit.

In case the expression is not in SNF or the semiring is not zero-divisor free, some edges are computed in  several times (coefficients are $\oplus$-added) which implies that some edges may be deleted. Then this characterization does not hold. A question then arises: the factorial condition is a sufficient condition to have an algorithm. Is it also a necessary condition ? 


{\small
\bibliographystyle{plain}
\bibliography{/Users/pacot/TEX/BIBLIO/biblio,/Users/pacot/TEX/BIBLIO/mar-bib}

\begin{thebibliography}{10}

\bibitem{BR88}
J.~Berstel and C.~Reutenauer.
\newblock {\em Rational series and their languages}.
\newblock EATCS Monographs on Theoretical Computer Science. Springer-Verlag,
  Berlin, 1988.

\bibitem{Bru93}
A.~Br{\"u}ggemann-Klein.
\newblock Regular expressions into finite automata.
\newblock {\em Theoret. Comput. Sci.}, 120(2):197--213, 1993.

\bibitem{BGW00}
A.~Buchsbaum, R.~Giancarlo, and J.~Westbrook.
\newblock On the determinization of weighted finite automata.
\newblock {\em SIAM J. Comput.}, 30(5):1502--1531, 2000.

\bibitem{CF03}
P.~Caron and M.~Flouret.
\newblock Glushkov construction for series: the non commutative case.
\newblock {\em Internat. J. Comput. Math.}, 80(4):457--472, 2003.

\bibitem{CZ97}
P.~Caron and D.~Ziadi.
\newblock Characterization of {G}lushkov automata.
\newblock {\em Theoret. Comput. Sci.}, 233(1--2):75--90, 2000.

\bibitem{Eil74}
S.~Eilenberg.
\newblock {\em Automata, languages and machines}, volume~A.
\newblock Academic Press, New York, 1974.

\bibitem{Glu60}
V.~M. Glushkov.
\newblock On a synthesis algorithm for abstract automata.
\newblock {\em Ukr. Matem. Zhurnal}, 12(2):147--156, 1960.
\newblock In Russian.

\bibitem{Glu61}
V.~M. Glushkov.
\newblock The abstract theory of automata.
\newblock {\em Russian Mathematical Surveys}, 16:1--53, 1961.

\bibitem{HW96}
U.~Hebisch and H.J. Weinert.
\newblock Semirings and semifields.
\newblock In M.~Hazewinkel, editor, {\em Handbook of Algebra}, volume~1,
  chapter~1F, pages 425--462. North-Holland, Amsterdam, 1996.

\bibitem{HU79}
J.~E. Hopcroft and J.~D. Ullman.
\newblock {\em Introduction to Automata Theory, Languages and Computation}.
\newblock Addison-Wesley, Reading, MA, 1979.

\bibitem{Kle56}
S.~Kleene.
\newblock Representation of events in nerve nets and finite automata.
\newblock {\em Automata Studies}, Ann. Math. Studies 34:3--41, 1956.
\newblock Princeton U. Press.

\bibitem{LS01}
S.~Lombardy and J.~Sakarovitch.
\newblock Derivatives of rational expressions with multiplicity.
\newblock {\em Theor. Comput. Sci.}, 332(1-3):141--177, 2005.

\bibitem{MY60}
R.~F. McNaughton and H.~Yamada.
\newblock Regular expressions and state graphs for automata.
\newblock {\em IEEE Transactions on Electronic Computers}, 9:39--57, March
  1960.

\bibitem{Sak03}
J.~Sakarovitch.
\newblock {\em \'El\'ements de th\'eorie des automates}.
\newblock Vuibert, Paris, 2003.

\bibitem{Sch61.1}
M.~P. Sch{\"u}tzenberger.
\newblock On the definition of a family of automata.
\newblock {\em Inform. and Control}, 4:245--270, 1961.

\end{thebibliography}
}
\end{document}